\documentclass{article}
\usepackage{amssymb}
\usepackage{graphicx}
\usepackage{hyperref}
\usepackage{listings}
\usepackage{mathtools}
\usepackage{maple}
\usepackage[utf8]{inputenc}
\usepackage[svgnames]{xcolor}
\usepackage{amsmath}
\usepackage{breqn}
\usepackage{textcomp}

\begin{document}
\lstset{basicstyle=\ttfamily,breaklines=true,columns=flexible}
\pagestyle{empty}
\DefineParaStyle{Maple Bullet Item}
\DefineParaStyle{Maple Heading 1}
\DefineParaStyle{Maple Warning}
\DefineParaStyle{Maple Heading 4}
\DefineParaStyle{Maple Heading 2}
\DefineParaStyle{Maple Heading 3}
\DefineParaStyle{Maple Dash Item}
\DefineParaStyle{Maple Error}
\DefineParaStyle{Maple Title}
\DefineParaStyle{Maple Ordered List 1}
\DefineParaStyle{Maple Text Output}
\DefineParaStyle{Maple Ordered List 2}
\DefineParaStyle{Maple Ordered List 3}
\DefineParaStyle{Maple Normal}
\DefineParaStyle{Maple Ordered List 4}
\DefineParaStyle{Maple Ordered List 5}
\DefineCharStyle{Maple 2D Output}
\DefineCharStyle{Maple 2D Input}
\DefineCharStyle{Maple Maple Input}
\DefineCharStyle{Maple 2D Math}
\DefineCharStyle{Maple Hyperlink}
\begin{center}
\begin{Maple Normal}

\end{Maple Normal}
\end{center}
\begin{center}
\begin{Maple Normal}
\textbf{Computer Algebra in Physics: }
\end{Maple Normal}
\end{center}
\begin{center}
\begin{Maple Normal}
\textbf{The hidden SO(4) symmetry of the hydrogen atom}
\end{Maple Normal}
\end{center}
\begin{center}
\begin{Maple Normal}

\end{Maple Normal}
\end{center}
\begin{center}
\begin{Maple Normal}
Pascal Szriftgiser\textsuperscript{1} and Edgardo S. Cheb-Terrab\textsuperscript{2} 
\end{Maple Normal}
\end{center}
\begin{center}
\begin{Maple Normal}
(1) University of Lille, CNRS, UMR 8523 - PhLAM - Physique des Lasers, Atomes et Molécules, F-59000 Lille, France
\end{Maple Normal}
\end{center}
\begin{center}
\begin{Maple Normal}
(2) Maplesoft, Canada
\end{Maple Normal}
\end{center}
\begin{center}
\begin{Maple Normal}

\end{Maple Normal}
\end{center}
\begin{Maple Normal}

\end{Maple Normal}
\begin{center}
\begin{Maple Normal}
\textbf{Abstract}
\end{Maple Normal}
\end{center}
\begin{center}
\begin{Maple Normal}

\end{Maple Normal}
\end{center}
\begin{Maple Normal}
Pauli first noticed the hidden SO(4) symmetry for the hydrogen atom in the early stages of quantum mechanics [1]. Starting from that symmetry, one can recover the spectrum of a spinless hydrogen atom and the degeneracy of its states without explicitly solving Schrödinger's equation [2]. In this paper, we derive that SO(4) symmetry and spectrum using a computer algebra system (CAS). While this problem is well known [3, 4], its solution involves several steps of manipulating expressions with tensorial quantum operators, including simplifying them by taking into account a combination of commutator rules and Einstein's sum rule for repeated indices. Therefore, it is an excellent model to test the current status of CAS concerning this kind of quantum-and-tensor-algebra computations and to showcase the CAS technique. Generally speaking, when capable, CAS can significantly help with manipulations that, like non-commutative tensor calculus subject to algebra rules, are tedious, time-consuming and error-prone. The presentation also shows two alternative patterns of computer algebra steps that can be used for systematically tackling more complicated symbolic problems of this kind.
\end{Maple Normal}
\begin{Maple Normal}

\end{Maple Normal}
\section*{\textbf{Introduction}}
\begin{Maple Normal}
In this work we derive, step-by-step, the SO(4) symmetry of the hydrogen atom and its spectrum using a computer algebra system (CAS). To the best of our knowledge, such a derivation using symbolic computation has not been shown before. The goal was to see whether this computation can be performed entering only the main definition formulas, followed by only simplification commands, and without using previous knowledge of the result. The presentation that follows showcases that approach, illustrating different techniques. The intricacy of this problem is in the symbolic manipulation and simplification of expressions involving noncommutative quantum tensor operators. The simplifications need to take into account commutator rules, symmetries under permutation of indices of tensorial subexpressions, and use Einstein's sum rule for repeated indices.
\end{Maple Normal}
\begin{Maple Normal}

\end{Maple Normal}
\begin{Maple Normal}
We performed the derivation using the Maple 2020 CAS with the \href{https://www.maplesoft.com/products/maple/features/physicsresearch.aspx}{Maplesoft Physics Updates v.705}. Generally speaking, the default computational domains of CAS don't include tensors, noncommutative operators nor related simplifications. The exception is the Maple system, which is distributed with the Physics package, which extends that default domain to include those objects and related operations. Relevant for our purpose, Physics includes a Simplify command which takes into account custom algebra rules and the sum rule for repeated indices, and uses tensor-simplification algorithms [5] adapted to work on a noncommutative domain.
\end{Maple Normal}
\begin{Maple Normal}

\end{Maple Normal}
\begin{Maple Normal}
A few notes about notation: when working with a CAS, besides the expectation of achieving a correct result for a complicated symbolic calculation, readability is also an issue. It is desired that one be able to enter the definition formulas and computational steps to be performed (the \textit{input}, shown in what follows preceded by a prompt 
{$ > $}) in a way that resembles as closely as possible their paper and pencil representation, and that the results (the \textit{output}, computed by the CAS) use easy-to-read, textbook mathematical-physics notation. The Maple Physics package implements such dedicated typesetting. In what follows, within text and in the \textit{output}, noncommutative objects are displayed using a different color, e.g. 
{$ \textcolor{olive}{H} $}, vectors and tensor indices are displayed in the standard way, as in 
{$ {\moverset{\rightarrow}{\textcolor{olive}{L}}} $} and 
{$ {\textcolor{olive}{L}{\idn{q}}} $}, and commutators are displayed with a minus subscript, e.g. 
{$ {\textcolor{gray}{[}\textcolor{olive}{H},{\textcolor{olive}{L}{\idn{q}}}\textcolor{gray}{]}}_{-} $}. Although the Maple system optionally provides dedicated typesetting also for the \textit{input}, we preferred to keep visible the Maple \textit{input} syntax, allowing for comparison with paper and pencil notation and to transmit a more accurate picture of what it is like to work on a real problem using CAS. We collected the names of the handful of commands used together with a one line description for them in an Appendix at the end. Maple also implements the concept of \textit{inert }representations of computations, which are activated only when desired. We use this feature in several places. Inert computations are entered by preceding the command with \% and are displayed in grey. Finally, as is usual in CAS, every output has an equation label, which we use throughout the presentation to refer to previous intermediate results, both in text and in input lines.
\end{Maple Normal}
\begin{Maple Normal}

\end{Maple Normal}
\begin{Maple Normal}
In Sec.1, we present the standard formulation of the problem and the computational goal, which is the derivation of the formulas representing the SO(4) symmetry and related spectrum.
\end{Maple Normal}
\begin{Maple Normal}

\end{Maple Normal}
\begin{Maple Normal}
In Sec.2, we formulate the problem on a Maple worksheet by setting tensorial non-commutative operators representing position, linear and angular momentum, respectively 
{$ {\textcolor{olive}{X}{\idn{a}}} $}, 
{$ {\textcolor{olive}{p}{\idn{a}}} $} and 
{$ \textcolor{olive}{L}_{a} $}, their commutation rules used as departure point, and the form of the quantum Hamiltonian 
{$ \textcolor{olive}{H} $}. That formulation is also used to derive a few related identities used in the sections that follow. 
\end{Maple Normal}
\begin{Maple Normal}

\end{Maple Normal}
\begin{Maple Normal}
In Sec.3, we derive the conservation of both angular momentum and the Runge-Lenz quantum operator, respectively 
{$ {\textcolor{gray}{[}\textcolor{olive}{H},{\textcolor{olive}{L}{\idn{q}}}\textcolor{gray}{]}}_{-} $} and 
{$ {\textcolor{gray}{[}\textcolor{olive}{H},{\textcolor{olive}{Z}{\idn{k}}}\textcolor{gray}{]}}_{-} $}. Taking advantage of the \textit{differentialoperators} feature of the Physics package, we explore two equivalent approaches: first, using only a symbolic tensor representation 
{$ \textcolor{olive}{p}_{j} $} of the momentum operator; second, using an explicit differential operator representation for it in configuration space, 
{$ \textcolor{olive}{p}_{j}=-i \hslash {\partial{\idn{j}}} $}. With the first approach, expressions are simplified only using the departing commutation rules and Einstein's sum rule for repeated indices. Using the second approach, 
{$ \textcolor{olive}{p}_{j} $} represents an abstract non-commutative differentiation operator with respect to the coordinates which acts over expressions that involve a test function 
{$ G (\textcolor{olive}{X}) $}. In the end, 
{$ \textcolor{olive}{p}_{j} $} is given an explicit form in coordinate representation the differentiation operations are performed and the test function 
{$ G (\textcolor{olive}{X}) $}is removed, yielding a result. Presenting both approaches is of interest as it offers two independent methods for performing the same computation, which is helpful to provide confidence in the results, a relevant issue when using computer algebra and in general.
\end{Maple Normal}
\begin{Maple Normal}

\end{Maple Normal}
\begin{Maple Normal}
In Sec.4, we derive 
{$ {{\textcolor{gray}{[}\textcolor{olive}{L}_{m},\textcolor{olive}{Z}_{n}\textcolor{gray}{]}}}_{-}=\mathrm{i}\hslash {\epsilon{\idn{m}\idn{n}\idn{u}}} \textcolor{olive}{Z}_{u} $} and show that the classical relation between angular momentum and the Runge-Lenz vectors,  
{$ {\moverset{\rightarrow}{\textcolor{olive}{L}}} $}
{$ . $}
{$ {\moverset{\rightarrow}{\textcolor{olive}{Z}}}=0 $}, due to the orbital momentum being perpendicular to the elliptic plane of motion in which the Runge-Lenz vector lies, still holds in quantum mechanics, where the components of these quantum vector operators do not commute.
\end{Maple Normal}
\begin{Maple Normal}

\end{Maple Normal}
\begin{Maple Normal}
In Sec.5, we derive 
{$ {\textcolor{gray}{[}{\textcolor{olive}{Z}{\idn{a}}},{\textcolor{olive}{Z}{\idn{b}}}\textcolor{gray}{]}}_{-}=-\frac{2 \mathrm{i} \hslash  {\epsilon{\idn{a}\idn{b}\idn{c}}}}{m_{e}}\textcolor{olive}{H} {\textcolor{olive}{L}{\idn{c}}} $} using the two alternative approaches described in Sec.3.
\end{Maple Normal}
\begin{Maple Normal}

\end{Maple Normal}
\begin{Maple Normal}
In Sec.6, we derive the well-known formula for the square of the Runge-Lenz vector, 
{$ {\textcolor{olive}{Z}{\idn{k}}}^{2}=\frac{2}{m_{e}}\textcolor{olive}{H} (\hslash^{2}+{\textcolor{olive}{L}{\idn{a}}}^{2})+\kappa^{2} $}.
\end{Maple Normal}
\begin{Maple Normal}

\end{Maple Normal}
\begin{Maple Normal}
In Sec.7, we use the SO(4) algebra derived in the previous sections to obtain the spectrum of the hydrogen atom. Following the literature, this approach is limited to the bound states for which the energy is negative. 
\end{Maple Normal}
\begin{Maple Normal}

\end{Maple Normal}
\begin{Maple Normal}
Some concluding remarks are presented at the end, and input syntax details are summarized in the Appendix. A Maple worksheet with the contents of this presentation, used to produce this article by exporting to LaTeX, can be downloaded from this Mapleprimes post: \href{https://www.mapleprimes.com/posts/208810-The-Hidden-SO4-Symmetry-Of-The-Hydrogen-Atom}{The-Hidden-SO4-Symmetry-Of-The-Hydrogen-Atom}.
\end{Maple Normal}
\section{\textbf{The hidden SO(4) symmetry of the hydrogen atom}}
\begin{Maple Normal}
Let's consider the hydrogen atom and its Hamiltonian
\end{Maple Normal}
\begin{center}
\begin{Maple Normal}
{$ \displaystyle H =\frac{{\|{\moverset{\rightarrow}{p}}\|}^{2}}{2m_{e}} -\frac{\kappa}{r} $},
\end{Maple Normal}
\end{center}
\begin{Maple Normal}
where 
{$ {\moverset{\rightarrow}{p}} $}is the electron momentum, 
{$ m_{e} $} its mass, 
{$ \kappa  $} a real positive constant, 
{$ r ={\|{\moverset{\rightarrow}{r}}\|}\esequiv \sqrt{{X{\idn{a}}}^{2}} $} the distance of the electron from the proton located at the origin, and 
{$ {X{\idn{a}}} $} is its tensorial representation with components 
{$ [x ,y ,z] $}. We assume that the proton's mass is infinite. The electron and nucleus spin are not taken into account. Classically, from the potential 
{$ -\frac{\kappa}{r} $}, one can derive a central force 
{$ {\moverset{\rightarrow}{F}}=-\kappa \frac{ \hat{r}}{r^{2}} $} that drives the electron's motion. Introducing the angular momentum
\end{Maple Normal}
\begin{center}
\begin{Maple Normal}
{$ {\moverset{\rightarrow}{L}}={\moverset{\rightarrow}{r}}\times {\moverset{\rightarrow}{p}} $},
\end{Maple Normal}
\end{center}
\begin{Maple Normal}
one can further define the Runge-Lenz vector 
{$ {\moverset{\rightarrow}{Z}}\colon  $}
\end{Maple Normal}
\begin{center}
\begin{Maple Normal}
{$ \displaystyle {\moverset{\rightarrow}{Z}}=\frac{1}{\mathit{m_{e}}}\cdot ({\moverset{\rightarrow}{L}}\times {\moverset{\rightarrow}{p}})+\kappa \frac{ {\moverset{\rightarrow}{r}}}{r} $}
\end{Maple Normal}
\end{center}
\begin{Maple Normal}
It is well known that 
{$ \frac{d}{d t}{\moverset{\rightarrow}{Z}}(t)=0 $}, 
{$ {\moverset{\rightarrow}{Z}} $} is a constant of the motion. Switching to Quantum Mechanics, this condition reads
\end{Maple Normal}
\begin{center}
\begin{Maple Normal}
{$ \displaystyle [\textcolor{olive}{H},{\moverset{\rightarrow}{\textcolor{olive}{Z}}}]_{-}=0 $}
\end{Maple Normal}
\end{center}
\begin{Maple Normal}
where, for hermiticity purpose, the expression of 
{$ {\moverset{\rightarrow}{\textcolor{olive}{Z}}} $} must be symmetrized
\end{Maple Normal}
\begin{center}
\begin{Maple Normal}
{$ \displaystyle {\moverset{\rightarrow}{\textcolor{olive}{Z}}}=\frac{1}{2\mathit{m_{e}}}({\moverset{\rightarrow}{\textcolor{olive}{L}}}\times {\moverset{\rightarrow}{\textcolor{olive}{p}}}-{\moverset{\rightarrow}{\textcolor{olive}{p}}}\times {\moverset{\rightarrow}{\textcolor{olive}{L}}})+\kappa \frac{{\moverset{\rightarrow}{\textcolor{olive}{r}}}}{r} $}
\end{Maple Normal}
\end{center}
\begin{Maple Normal}
In what follows, departing from the Hamiltonian \textit{H}\textit{,} the basic commutation rules between position 
{$ {\moverset{\rightarrow}{\textcolor{olive}{r}}} $}, momentum 
{$ {\moverset{\rightarrow}{\textcolor{olive}{p}}} $} and angular momentum 
{$ {\moverset{\rightarrow}{\textcolor{olive}{L}}} $} in tensor notation, we derive the following commutation rules between the quantum Hamiltonian, angular momentum and Runge-Lenz vector 
{$ {\moverset{\rightarrow}{\textcolor{olive}{Z}}} $}
\end{Maple Normal}
\begin{center}
\begin{tabular}{ r c l }
\vspace{1mm}
{$ \displaystyle [\textcolor{olive}{H},\textcolor{olive}{L}_{n}]_{-} $} & = & 0
\\ \vspace{1mm}
{$ \displaystyle [\textcolor{olive}{H},\textcolor{olive}{Z}_{n}]_{-} $} & = & 0
\\ \vspace{1mm}
{$ \displaystyle [\textcolor{olive}{L}_{m},\textcolor{olive}{Z}_{n}]_{-} $} & = & {$ \displaystyle \mathrm{i}\hslash  {\epsilon{\idn{m}\idn{n}\idn{o}}} \textcolor{olive}{Z}_{o} $}
\\ \vspace{1mm}
{$ \displaystyle [\textcolor{olive}{Z}_{m},\textcolor{olive}{Z}_{n}]_{-} $} & = & {$ \displaystyle -2\frac{\mathrm{i}\hslash}{m_{e}} \textcolor{olive}{H}{\epsilon{\idn{m}\idn{n}\idn{o}}} \textcolor{olive}{L}_{o} $}
\end{tabular}
\end{center}
\begin{Maple Normal}
Since \textit{H} commutes with both 
{$ {\moverset{\rightarrow}{\textcolor{olive}{L}}} $} and 
{$ {\moverset{\rightarrow}{\textcolor{olive}{Z}}} $}, defining
\end{Maple Normal}
\begin{center}
\begin{Maple Normal}
{$ \displaystyle \textcolor{olive}{M}_{n}=\sqrt{-\frac{\mathit{m_{e}}}{2 \textcolor{olive}{H}}}\, \textcolor{olive}{Z}_{n}, $}
\end{Maple Normal}
\end{center}
\begin{Maple Normal}
these commutation rules can be rewritten as
\end{Maple Normal}
\begin{center}
\begin{tabular}{ r c l }
\vspace{1mm}
{$ \displaystyle [\textcolor{olive}{L}_{m},\textcolor{olive}{L}_{n}]_{-} $} & = & {$ \displaystyle \mathrm{i}\hslash  {\epsilon{\idn{m}\idn{n}\idn{o}}} \textcolor{olive}{L}_{o} $}
\\ \vspace{1mm}
{$ \displaystyle [\textcolor{olive}{L}_{m},\textcolor{olive}{M}_{n}]_{-} $} & = & {$ \displaystyle \mathrm{i}\hslash  {\epsilon{\idn{m}\idn{n}\idn{o}}} \textcolor{olive}{M}_{o} $}
\\ \vspace{1mm}
{$ \displaystyle [\textcolor{olive}{M}_{m},\textcolor{olive}{M}_{n}]_{-} $} & = & {$ \displaystyle \mathrm{i}\hslash {\epsilon{\idn{m}\idn{n}\idn{o}}} \textcolor{olive}{L}_{o} $}
\end{tabular}
\end{center}
\begin{Maple Normal}
This set constitutes the Lie algebra of the SO(4) group.
\end{Maple Normal}
\section{\textbf{Setting the problem, commutation rules and useful identities}}
\begin{Maple Normal}
Formulating the problem requires loading the \textit{Physics} package and its \textit{Library} and \textit{Vectors} subpackages, that contain additional manipulation commands; we set the imaginary unit to be represented by a lowercase Latin i letter instead of the default uppercase I.
\end{Maple Normal}
\mapleinput
{$ \displaystyle \texttt{>\,} \mathit{with} (\mathit{Physics})\colon 
 \mathit{with} (\mathit{Library})\colon 
 \mathit{with} (\mathit{Vectors})\colon 
 \mathit{interface} (\mathit{imaginaryunit} =\,i)\colon  $}

\begin{Maple Normal}

\end{Maple Normal}
\begin{Maple Normal}
The context for this problem is Cartesian coordinates and a 3D Euclidean space where all of 
{$ {\{\hslash ,\kappa ,m_{e}\}} $} are real objects. We chose lowercase letters to represent tensor indices, the use of automatic simplification (i.e., \textit{automatically simplify the size} of everything being displayed)
\end{Maple Normal}
\mapleinput
{$ \displaystyle \texttt{>\,} \mathit{Setup} (\mathit{coordinates} =\,\mathit{cartesian} ,\mathit{realobjects} =\{\hslash ,\kappa ,m_{e}\},
\\
\mathit{automaticsimplification} =\mathit{true} ,\mathit{dimension} =\,3,\mathit{metric} =\mathit{Euclidean} ,
\\
\mathit{spacetimeindices} =\,\mathit{lowercaselatin} ,\mathit{quiet}) $}

\begin{dmath}\label{(1)}
\left[\mathit{automaticsimplification}  \hiderel{=} \mathit{true} ,\mathit{coordinatesystems}  \hiderel{=} \left\{X \right\},
\\
\mathit{dimension} \hiderel{=}3,\mathit{metric}  \hiderel{=} \left\{\left(1,1\right) \hiderel{=} 1,\left(2,2\right) \hiderel{=} 1,\left(3,3\right) \hiderel{=} 1\right\},
\\
\mathit{realobjects} \hiderel{=}\left\{\hslash ,\kappa ,m_{e},\phi ,r ,\rho ,\theta ,x ,y ,z \right\},\mathit{spacetimeindices}  \hiderel{=} \mathit{lowercaselatin} \right]
\end{dmath}
\begin{Maple Normal}
Next, we set the quantum Hermitian operators (not \textit{Z}, we derive that property for it further below) and related commutators:
\end{Maple Normal}
\begin{Maple Dash Item}
\item the dimensionless potential  
{$ \textcolor{olive}{V}=\frac{1}{\textcolor{olive}{r}} $} is assumed to commute with position, not with momentum - the commutation rule with 
{$ \textcolor{olive}{p}_{k} $} is derived in Sec.3.2;
\item the commutator rules between position 
{$ {\textcolor{olive}{X}{\idn{n}}} $} on the one hand, and linear 
{$ \textcolor{olive}{p}_{k} $} and angular momentum 
{$ \textcolor{olive}{L}_{k} $} on the other hand, are the departure point, entered using the inert form of the \textit{Commutator} command. Tensors are indexed using the standard Maple notation for indexation, [].
\end{Maple Dash Item}
\mapleinput
{$ \displaystyle \texttt{>\,} \mathit{Setup} (\mathit{quantumoperators} =\{Z \}, \mathit{hermitianoperators} =\{V ,H ,L ,X ,p \},
\\
\mathit{algebrarules} =\{\mathit{\%\!Commutator} (p [k],p [l])=0,
\\
\mathit{\%\!Commutator} (X [k],p [l])=\mathrm{i}\, \hslash \, \textit{g\_} [k ,l],
\\
\mathit{\%\!Commutator} (L [j],L [k])=\mathrm{i}\, \hslash \, \mathit{LeviCivita} [j ,k ,n]\, L [n],
\\
\mathit{\%\!Commutator} (p [j],L [k])=\mathrm{i}\, \hslash \, \mathit{LeviCivita} [j ,k ,n]\, p [n],
\\
\mathit{\%\!Commutator} (X [j],L [k])=\mathrm{i}\, \hslash \, \mathit{LeviCivita} [j ,k ,n]\, X [n],
\\
\mathit{\%\!Commutator} (X [k],V (X))=0\}) $}

\begin{dmath}\label{(2)}
\left[\mathit{algebrarules}  \hiderel{=} \left\{\left[\textcolor{olive}{L}_{j},\textcolor{olive}{L}_{k}\right]_{-} \hiderel{=} \mathrm{i} \hslash  {\epsilon{\idn{j}\idn{k}\idn{n}}} \textcolor{olive}{L}_{n},\left[\textcolor{olive}{p}_{j},\textcolor{olive}{L}_{k}\right]_{-} \hiderel{=} \mathrm{i} \hslash  {\epsilon{\idn{j}\idn{k}\idn{n}}} \textcolor{olive}{p}_{n},\left[\textcolor{olive}{p}_{k},\textcolor{olive}{p}_{l}\right]_{-} \hiderel{=} 0,
\\
\left[{\textcolor{olive}{X}{\idn{j}}},\textcolor{olive}{L}_{k}\right]_{-}\hiderel{=}\mathrm{i} \hslash  {\epsilon{\idn{j}\idn{k}\idn{n}}} {\textcolor{olive}{X}{\idn{n}}},\left[{\textcolor{olive}{X}{\idn{k}}},\textcolor{olive}{p}_{l}\right]_{-} \hiderel{=} \mathrm{i} \hslash  {g{\idn{k}\idn{l}}},\left[{\textcolor{olive}{X}{\idn{k}}},\textcolor{olive}{V}\! \left(\textcolor{olive}{X}\right)\right]_{-} \hiderel{=} 0\right\},
\\
\mathit{hermitianoperators} \hiderel{=}\left\{\textcolor{olive}{H},\textcolor{olive}{L},\textcolor{olive}{V},\textcolor{olive}{p},\textcolor{olive}{x},\textcolor{olive}{y},\textcolor{olive}{z}\right\},
\\
\mathit{quantumoperators} \hiderel{=}\left\{\textcolor{olive}{H},\textcolor{olive}{L},\textcolor{olive}{V},\textcolor{olive}{Z},\textcolor{olive}{p},\textcolor{olive}{x},\textcolor{olive}{y},\textcolor{olive}{z}\right\}\right]
\end{dmath}
\begin{Maple Normal}
Define the \textit{tensor} quantum operators representing the linear momentum, angular momentum and the Runge-Lenz vectors
\end{Maple Normal}
\mapleinput
{$ \displaystyle \texttt{>\,} \mathit{Define} (p [k],L [k],Z [k],\mathit{quiet}) $}

\begin{dmath}\label{(3)}
\left\{{\textcolor{olive}{\gamma}{\idn{a}}},{\textcolor{olive}{L}{\idn{k}}},{\textcolor{olive}{\sigma}{\idn{a}}},{\textcolor{olive}{Z}{\idn{k}}},{\partial{\idn{a}}},{g{\idn{a}\idn{b}}},{\textcolor{olive}{p}{\idn{k}}},{\epsilon{\idn{a}\idn{b}\idn{c}}},{\textcolor{olive}{X}{\idn{a}}}\right\}
\end{dmath}
\begin{Maple Normal}
For readability, avoid redundant display of functionality
\end{Maple Normal}
\mapleinput
{$ \displaystyle \texttt{>\,} \mathit{CompactDisplay} ((V ,G)(X)) $}

\begin{maplelatex}
\mapleinline{inert}{2d}{}
{\[\textcolor{olive}{V}(\textcolor{olive}{X}) \mathit{will\,now\,be\,displayed\,as\,} \textcolor{olive}{V}\]}
\end{maplelatex}
\begin{dmath}\label{(4)}
G \! \left(\textcolor{olive}{X}\right) \mathit{will\,now\,be\,displayed\,as\,} G 
\end{dmath}
\begin{Maple Normal}
The Hamiltonian for the hydrogen atom is entered as
\end{Maple Normal}
\mapleinput
{$ \displaystyle \texttt{>\,} H =\frac{p [l]^{2}}{2\mathit{\cdot\,}\mathit{m_{e}}}-\kappa \cdot V (X) $}

\begin{dmath}\label{(5)}
\textcolor{olive}{H}=\frac{{\textcolor{olive}{p}{\idn{l}\iup{2}}}}{2 m_{e}}-\kappa  \textcolor{olive}{V}
\end{dmath}
\subsection{\textbf{Definition of V(X) and related identities}}
\begin{Maple Normal}
We use the dimensionless potential 
{$ \textcolor{olive}{V}(\textcolor{olive}{X}) $}
\end{Maple Normal}
\mapleinput
{$ \displaystyle \texttt{>\,} V (X)=\frac{1}{(X [l]^{2})^{\frac{1}{2}}} $}

\begin{dmath}\label{(6)}
\textcolor{olive}{V}=\left({\textcolor{olive}{X}{\idn{l}\iup{2}}}\right)^{-\frac{1}{2}}
\end{dmath}
\begin{Maple Normal}
The gradient of 
{$ \textcolor{olive}{V}(\textcolor{olive}{X}) $} is 
\end{Maple Normal}
\mapleinput
{$ \displaystyle \texttt{>\,} \textit{d\_} [n](\mapleref{(6)}) $}

\begin{dmath}\label{(7)}
{\partial{\idn{n}}}\! \left(\textcolor{olive}{V}\right)=-\left({\textcolor{olive}{X}{\idn{l}\iup{2}}}\right)^{-\frac{3}{2}} {\textcolor{olive}{X}{\idn{n}}}
\end{dmath}
\begin{Maple Normal}
where we note that all these commands (including product and power), \textit{distribute over equations}. So that
\end{Maple Normal}
\mapleinput
{$ \displaystyle \texttt{>\,} \mathit{subs} ((\mathit{rhs} =\mathit{lhs})(\mapleref{(6)}^{3}),\mapleref{(7)}) $}

\begin{dmath}\label{(8)}
{\partial{\idn{n}}}\! \left(\textcolor{olive}{V}\right)=-{\textcolor{olive}{V}}^{3} {\textcolor{olive}{X}{\idn{n}}}
\end{dmath}
\begin{Maple Normal}
Equivalently, from \mapleref{(6)} one can deduce  that will often be used afterwards
\end{Maple Normal}
\mapleinput
{$ \displaystyle \texttt{>\,} (\mathit{rhs} =\mathit{lhs})(\frac{V (X)^{3}}{\mapleref{(6)}^{2}}) $}

\begin{dmath}\label{(9)}
{\textcolor{olive}{V}}^{3} {\textcolor{olive}{X}{\idn{l}\iup{2}}}=\textcolor{olive}{V}
\end{dmath}
\subsection{\textbf{The commutation rules between linear and angular momentum and of the potential V(X)}}
\begin{Maple Normal}
By definition
\end{Maple Normal}
\mapleinput
{$ \displaystyle \texttt{>\,} L [q]=\,\mathit{LeviCivita} [q ,m ,n]\cdot X [m]\,\cdot p [n] $}

\begin{dmath}\label{(10)}
{\textcolor{olive}{L}{\idn{q}}}={\epsilon{\idn{m}\idn{n}\idn{q}}} {\textcolor{olive}{X}{\idn{m}}} {\textcolor{olive}{p}{\idn{n}}}
\end{dmath}
\begin{Maple Normal}
so
\end{Maple Normal}
\mapleinput
{$ \displaystyle \texttt{>\,} \mathit{Commutator} (\mapleref{(10)},V (X)) $}

\begin{dmath}\label{(11)}
\left[{\textcolor{olive}{L}{\idn{q}}},\textcolor{olive}{V}\right]_{-}={\epsilon{\idn{m}\idn{n}\idn{q}}} {\textcolor{olive}{X}{\idn{m}}} \left[{\textcolor{olive}{p}{\idn{n}}},\textcolor{olive}{V}\right]_{-}
\end{dmath}
\begin{Maple Normal}
The commutator on the right-hand side cannot be computed by the CAS until providing more information. To derive the value of 
{$ [{\textcolor{olive}{p}{\idn{n}}},\textcolor{olive}{V}]_{-} $}we set 
{$ {\textcolor{olive}{p}{\idn{n}}} $} as a \textit{differentialoperator} and introduce an arbitrary test function 
{$ G (\textcolor{olive}{X}) $}
\end{Maple Normal}
\mapleinput
{$ \displaystyle \texttt{>\,} \mathit{Setup} (\mathit{differentialoperators} =\{[p [k],[x ,y ,z]]\}) $}

\begin{dmath}\label{(12)}
\left[\mathit{differentialoperators} =\left\{\left[{\textcolor{olive}{p}{\idn{k}}},\left[\textcolor{olive}{X}\right]\right]\right\}\right]
\end{dmath}
\begin{Maple Normal}
Apply now to 
{$ G (\textcolor{olive}{X}) $} the differential operator 
{$ {\textcolor{olive}{p}{\idn{n}}} $} found in the commutator of the right-hand side of \mapleref{(11)} 
\end{Maple Normal}
\mapleinput
{$ \displaystyle \texttt{>\,} (\mathit{lhs} =\mathit{ApplyProductsOfDifferentialOperators} @ \mathit{rhs})(\mapleref{(11)}\cdot G (X)) $}

\begin{dmath}\label{(13)}
\left[{\textcolor{olive}{L}{\idn{q}}},\textcolor{olive}{V}\right]_{-} G ={\epsilon{\idn{m}\idn{n}\idn{q}}} {\textcolor{olive}{X}{\idn{m}}} \left({\textcolor{olive}{p}{\idn{n}}}\! \left(\textcolor{olive}{V} G \right)-\textcolor{olive}{V} {\textcolor{olive}{p}{\idn{n}}}\! \left(G \right)\right)
\end{dmath}
\begin{Maple Normal}
The result of 
{$ {\textcolor{olive}{p}{\idn{l}}}(G (\textcolor{olive}{X})) $} is not known to the system at this point. Define then an explicit representation for 
{$ \textcolor{olive}{p}_{n} $} as the differential operator in configuration space 
{$ \textcolor{olive}{p}_{n}=-i \hslash {\partial{\idn{n}}} $}
\end{Maple Normal}
\mapleinput
{$ \displaystyle \texttt{>\,} p \coloneqq u \,\rightarrow -i \,\hslash \cdot \textit{d\_} [\mathit{op} (\mathit{procname})](u) $}

\begin{dmath}\label{(14)}
\textcolor{olive}{p}\coloneqq u \hiderel{\mapsto }\mathrm{-i} \hslash  {\partial{\idn{\mathit{op}}\idn{\!}\idn{\left(\mathit{procname} \right)}}}\! \left(u \right)
\end{dmath}
\begin{Maple Normal}
where in the above 
{$ \mathit{op} (\mathit{procname}) $} represents the indices with which the differential operator 
{$ \textcolor{olive}{p}_{n} $} is called.With this definition, the right-hand side of \mapleref{(13)}  automatically evaluates to
\end{Maple Normal}
\mapleinput
{$ \displaystyle \texttt{>\,} \mapleref{(13)} $}

\begin{dmath}\label{(15)}
\left[{\textcolor{olive}{L}{\idn{q}}},\textcolor{olive}{V}\right]_{-} G =\mathrm{-i} {\epsilon{\idn{m}\idn{n}\idn{q}}} \hslash  {\textcolor{olive}{X}{\idn{m}}} {\partial{\idn{n}}}\! \left(\textcolor{olive}{V}\right) G 
\end{dmath}
\begin{Maple Normal}
So that using \mapleref{(8)}
{$ \esequiv {\partial{\idn{n}}}(\textcolor{olive}{V})=-{\textcolor{olive}{V}}^{3} {\textcolor{olive}{X}{\idn{n}}} $} and multiplying by 
{$ {\textcolor{olive}{G}(\textcolor{olive}{X})}^{\mathit{-}1} $}, 
\end{Maple Normal}
\mapleinput
{$ \displaystyle \texttt{>\,} \mathit{SubstituteTensor} (\mapleref{(8)},\mapleref{(15)})\cdot G (X)^{\mathit{-}1} $}

\begin{dmath}\label{(16)}
\left[{\textcolor{olive}{L}{\idn{q}}},\textcolor{olive}{V}\right]_{-}=\mathrm{i} {\epsilon{\idn{m}\idn{n}\idn{q}}} \hslash  {\textcolor{olive}{X}{\idn{m}}} {\textcolor{olive}{V}}^{3} {\textcolor{olive}{X}{\idn{n}}}
\end{dmath}
\begin{Maple Normal}
from where we get the first commutation rule:
\end{Maple Normal}
\mapleinput
{$ \displaystyle \texttt{>\,} \mathit{Simplify} (\mapleref{(16)}) $}

\begin{dmath}\label{(17)}
\left[{\textcolor{olive}{L}{\idn{q}}},\textcolor{olive}{V}\right]_{-}=0
\end{dmath}
\begin{Maple Normal}
Likewise, from the 
{$ \mathit{inert} =\mathit{active}  $} form of 
{$ [{\textcolor{olive}{p}{\idn{q}}},V (X)]_{-} $}
\end{Maple Normal}
\mapleinput
{$ \displaystyle \texttt{>\,} (\mathit{\%\!Commutator} =\mathit{Commutator})(p [q],V (X)) $}

\begin{dmath}\label{(18)}
\left[{\textcolor{olive}{p}{\idn{q}}},\textcolor{olive}{V}\right]_{-}=\left[{\textcolor{olive}{p}{\idn{q}}},\textcolor{olive}{V}\right]_{-}
\end{dmath}
\begin{Maple Normal}
by applying this equation to the test function 
{$ \textcolor{olive}{G}(\textcolor{olive}{X}) $} we get
\end{Maple Normal}
\mapleinput
{$ \displaystyle \texttt{>\,} (\mathit{lhs} =\mathit{ApplyProductsOfDifferentialOperators} @ \mathit{rhs})(\mapleref{(18)}\cdot G (X)) $}

\begin{dmath}\label{(19)}
\left[{\textcolor{olive}{p}{\idn{q}}},\textcolor{olive}{V}\right]_{-} G =\mathrm{-i} \hslash  {\partial{\idn{q}}}\! \left(\textcolor{olive}{V}\right) G 
\end{dmath}
\mapleinput
{$ \displaystyle \texttt{>\,} \mathit{SubstituteTensor} (\mapleref{(8)},\mapleref{(19)})\cdot G (X)^{\mathit{-}1} $}

\begin{dmath}\label{(20)}
\left[{\textcolor{olive}{p}{\idn{q}}},\textcolor{olive}{V}\right]_{-}=\mathrm{i} \hslash  {\textcolor{olive}{V}}^{3} {\textcolor{olive}{X}{\idn{q}}}
\end{dmath}
\begin{Maple Normal}
In the same way, for 
{$ {\textcolor{gray}{[}{\textcolor{olive}{p}{\idn{q}}},{\textcolor{olive}{V}}^{3}\textcolor{gray}{]}}_{-} $}  we get
\end{Maple Normal}
\mapleinput
{$ \displaystyle \texttt{>\,} (\mathit{\%\!Commutator} =\mathit{Commutator})(p [q],V (X)^{3}) $}

\begin{dmath}\label{(21)}
\left[{\textcolor{olive}{p}{\idn{q}}},{\textcolor{olive}{V}}^{3}\right]_{-}=\left[{\textcolor{olive}{p}{\idn{q}}},{\textcolor{olive}{V}}^{3}\right]_{-}
\end{dmath}
\mapleinput
{$ \displaystyle \texttt{>\,} (\mathit{lhs} =\mathit{ApplyProductsOfDifferentialOperators} @ \mathit{rhs})(\mapleref{(21)}\cdot G (X)) $}

\begin{dmath}\label{(22)}
\left[{\textcolor{olive}{p}{\idn{q}}},{\textcolor{olive}{V}}^{3}\right]_{-} G =\mathrm{-i} \hslash  \left({\partial{\idn{q}}}\! \left(\textcolor{olive}{V}\right) {\textcolor{olive}{V}}^{2}+\textcolor{olive}{V} {\partial{\idn{q}}}\! \left(\textcolor{olive}{V}\right) \textcolor{olive}{V}+{\textcolor{olive}{V}}^{2} {\partial{\idn{q}}}\! \left(\textcolor{olive}{V}\right)\right) G 
\end{dmath}
\mapleinput
{$ \displaystyle \texttt{>\,} \mathit{SubstituteTensor} (\mapleref{(8)},\mapleref{(22)})\cdot G (X)^{-1} $}

\begin{dmath}\label{(23)}
\left[{\textcolor{olive}{p}{\idn{q}}},{\textcolor{olive}{V}}^{3}\right]_{-}=\mathrm{i} \hslash  \left({\textcolor{olive}{V}}^{3} {\textcolor{olive}{X}{\idn{q}}} {\textcolor{olive}{V}}^{2}+{\textcolor{olive}{V}}^{4} {\textcolor{olive}{X}{\idn{q}}} \textcolor{olive}{V}+{\textcolor{olive}{V}}^{5} {\textcolor{olive}{X}{\idn{q}}}\right)
\end{dmath}
\mapleinput
{$ \displaystyle \texttt{>\,} (\mathit{lhs} =\mathit{Simplify} @ \mathit{rhs})(\mapleref{(23)}) $}

\begin{dmath}\label{(24)}
\left[{\textcolor{olive}{p}{\idn{q}}},{\textcolor{olive}{V}}^{3}\right]_{-}=3 \mathrm{i} \hslash  {\textcolor{olive}{V}}^{5} {\textcolor{olive}{X}{\idn{q}}}
\end{dmath}
\begin{Maple Normal}
Adding now these new commutation rules to the setup of the problem, they will be taken into account in subsequent uses of \textit{Simplify}
\end{Maple Normal}
\mapleinput
{$ \displaystyle \texttt{>\,} \mapleref{(17)},\mapleref{(20)},\mapleref{(24)} $}

\begin{dmath}\label{(25)}
\left[{\textcolor{olive}{L}{\idn{q}}},\textcolor{olive}{V}\right]_{-} \hiderel{=} 0,\left[{\textcolor{olive}{p}{\idn{q}}},\textcolor{olive}{V}\right]_{-} \hiderel{=} \mathrm{i} \hslash  {\textcolor{olive}{V}}^{3} {\textcolor{olive}{X}{\idn{q}}},\left[{\textcolor{olive}{p}{\idn{q}}},{\textcolor{olive}{V}}^{3}\right]_{-} \hiderel{=} 3 \mathrm{i} \hslash  {\textcolor{olive}{V}}^{5} {\textcolor{olive}{X}{\idn{q}}}
\end{dmath}
\mapleinput
{$ \displaystyle \texttt{>\,} \mathit{Setup} (\mapleref{(25)}) $}

\begin{dmath}\label{(26)}
\left[\mathit{algebrarules} =\left\{\left[{\textcolor{olive}{L}{\idn{j}}},{\textcolor{olive}{L}{\idn{k}}}\right]_{-} \hiderel{=} \mathrm{i} \hslash  {\epsilon{\idn{j}\idn{k}\idn{n}}} {\textcolor{olive}{L}{\idn{n}}},\left[{\textcolor{olive}{L}{\idn{q}}},\textcolor{olive}{V}\right]_{-} \hiderel{=} 0,\left[{\textcolor{olive}{p}{\idn{j}}},{\textcolor{olive}{L}{\idn{k}}}\right]_{-} \hiderel{=} \mathrm{i} \hslash  {\epsilon{\idn{j}\idn{k}\idn{n}}} {\textcolor{olive}{p}{\idn{n}}},
\\
\left[{\textcolor{olive}{p}{\idn{k}}},{\textcolor{olive}{p}{\idn{l}}}\right]_{-}\hiderel{=}0,\left[{\textcolor{olive}{p}{\idn{q}}},\textcolor{olive}{V}\right]_{-} \hiderel{=} \mathrm{i} \hslash  {\textcolor{olive}{V}}^{3} {\textcolor{olive}{X}{\idn{q}}},\left[{\textcolor{olive}{p}{\idn{q}}},{\textcolor{olive}{V}}^{3}\right]_{-} \hiderel{=} 3 \mathrm{i} \hslash  {\textcolor{olive}{V}}^{5} {\textcolor{olive}{X}{\idn{q}}},
\\
\left[{\textcolor{olive}{X}{\idn{j}}},{\textcolor{olive}{L}{\idn{k}}}\right]_{-}\hiderel{=}\mathrm{i} \hslash  {\epsilon{\idn{j}\idn{k}\idn{n}}} {\textcolor{olive}{X}{\idn{n}}},\left[{\textcolor{olive}{X}{\idn{k}}},{\textcolor{olive}{p}{\idn{l}}}\right]_{-} \hiderel{=} \mathrm{i} \hslash  {g{\idn{k}\idn{l}}},\left[{\textcolor{olive}{X}{\idn{k}}},\textcolor{olive}{V}\right]_{-} \hiderel{=} 0\right\}\right]
\end{dmath}
\begin{Maple Normal}
Undo \textit{differentialoperators} to work using two different approaches, with and without them.
\end{Maple Normal}
\mapleinput
{$ \displaystyle \texttt{>\,} \mathit{Setup} (\mathit{differentialoperators} =\mathit{none}) $}

\begin{dmath}\label{(27)}
\left[\mathit{differentialoperators} =\mathit{none} \right]
\end{dmath}
\section{\textbf{Commutation rules between the Hamiltonian and each of the angular momentum and Runge-Lenz tensors}}
\begin{Maple Normal}
Departing from the Hamiltonian  \mapleref{(5)}
{$ \esequiv \textcolor{olive}{H}=\frac{ {\textcolor{olive}{p}{\idn{l}}}^{2}}{2m_{e}}-\kappa  \textcolor{olive}{V} $} and the definition of angular momentum \mapleref{(10)}
{$ \esequiv {\textcolor{olive}{L}{\idn{q}}}={\epsilon{\idn{m}\idn{n}\idn{q}}} {\textcolor{olive}{X}{\idn{m}}} {\textcolor{olive}{p}{\idn{n}}} $}, by taking their commutator we get
\end{Maple Normal}
\mapleinput
{$ \displaystyle \texttt{>\,} \mathit{Commutator} (\mapleref{(5)},\mapleref{(10)}) $}

\begin{dmath}\label{(28)}
\left[\textcolor{olive}{H},{\textcolor{olive}{L}{\idn{q}}}\right]_{-}=\frac{\mathrm{-i} {\epsilon{\idn{m}\idn{n}\idn{q}}} \hslash  \left(-{\textcolor{olive}{X}{\idn{m}}} {\textcolor{olive}{V}}^{3} {\textcolor{olive}{X}{\idn{n}}} \kappa  m_{e}+{\textcolor{olive}{p}{\idn{l}}} {\textcolor{olive}{p}{\idn{n}}} {g{\idn{l}\idn{m}}}\right)}{m_{e}}
\end{dmath}
\mapleinput
{$ \displaystyle \texttt{>\,} \mathit{Simplify} (\mapleref{(28)}) $}

\begin{dmath}\label{(29)}
\left[\textcolor{olive}{H},{\textcolor{olive}{L}{\idn{q}}}\right]_{-}=0
\end{dmath}
\subsection{\textbf{The commutator between the Hamiltonian and Runge-Lenz tensor: algebraic approach}}
\begin{Maple Normal}
Start from the definition of the quantum Runge-Lenz tensor
\end{Maple Normal}
\mapleinput
{$ \displaystyle \texttt{>\,} Z [k]=\frac{1}{2\,\mathit{m_{e}}}\cdot \mathit{LeviCivita} [a ,b ,k]\cdot (L [a]\cdot p [b]-p [a]\cdot L [b])+\kappa \cdot V (X)\cdot X [k] $}

\begin{dmath}\label{(30)}
{\textcolor{olive}{Z}{\idn{k}}}=\frac{{\epsilon{\idn{a}\idn{b}\idn{k}}} \left({\textcolor{olive}{L}{\idn{a}}} {\textcolor{olive}{p}{\idn{b}}}-{\textcolor{olive}{p}{\idn{a}}} {\textcolor{olive}{L}{\idn{b}}}\right)}{2 m_{e}}+\kappa  \textcolor{olive}{V} {\textcolor{olive}{X}{\idn{k}}}
\end{dmath}
\begin{Maple Normal}
This tensor is Hermitian
\end{Maple Normal}
\mapleinput
{$ \displaystyle \texttt{>\,} \mapleref{(30)}-\mathit{Dagger} (\mapleref{(30)}) $}

\begin{dmath}\label{(31)}
{\textcolor{olive}{Z}{\idn{k}}}-{\textcolor{olive}{Z}{\idn{k}}}^{\dag}=\frac{2 \kappa  \textcolor{olive}{V} {\textcolor{olive}{X}{\idn{k}}} m_{e}-2 \kappa  {\textcolor{olive}{X}{\idn{k}}} \textcolor{olive}{V} m_{e}+{\epsilon{\idn{a}\idn{b}\idn{k}}} \left({\textcolor{olive}{L}{\idn{a}}} {\textcolor{olive}{p}{\idn{b}}}+{\textcolor{olive}{L}{\idn{b}}} {\textcolor{olive}{p}{\idn{a}}}-{\textcolor{olive}{p}{\idn{a}}} {\textcolor{olive}{L}{\idn{b}}}-{\textcolor{olive}{p}{\idn{b}}} {\textcolor{olive}{L}{\idn{a}}}\right)}{2 m_{e}}
\end{dmath}
\mapleinput
{$ \displaystyle \texttt{>\,} \mathit{Simplify} (\mapleref{(31)}) $}

\begin{dmath}\label{(32)}
{\textcolor{olive}{Z}{\idn{k}}}-{\textcolor{olive}{Z}{\idn{k}}}^{\dag}=0
\end{dmath}
\begin{Maple Normal}
Since the system knows about the commutation rule between linear and angular momentum,
\end{Maple Normal}
\mapleinput
{$ \displaystyle \texttt{>\,} (\mathit{\%\!Commutator} =\,\mathit{Commutator})(L [a],p [b]) $}

\begin{dmath}\label{(33)}
\left[{\textcolor{olive}{L}{\idn{a}}},{\textcolor{olive}{p}{\idn{b}}}\right]_{-}=\mathrm{i} \hslash  {\epsilon{\idn{a}\idn{b}\idn{n}}} {\textcolor{olive}{p}{\idn{n}}}
\end{dmath}
\begin{Maple Normal}
the expression \mapleref{(30)} for 
{$ {\textcolor{olive}{Z}{\idn{k}}} $} can be simplified
\end{Maple Normal}
\mapleinput
{$ \displaystyle \texttt{>\,} \mathit{Simplify} (\mapleref{(30)}) $}

\begin{dmath}\label{(34)}
{\textcolor{olive}{Z}{\idn{k}}}=\frac{\mathrm{i} \hslash  {\textcolor{olive}{p}{\idn{k}}}}{m_{e}}+\kappa  \textcolor{olive}{V} {\textcolor{olive}{X}{\idn{k}}}-\frac{{\epsilon{\idn{a}\idn{b}\idn{k}}} {\textcolor{olive}{p}{\idn{a}}} {\textcolor{olive}{L}{\idn{b}}}}{m_{e}}
\end{dmath}
\begin{Maple Normal}
and the angular momentum removed from the the right-hand side using \mapleref{(10)}
{$ \esequiv {\textcolor{olive}{L}{\idn{q}}}={\epsilon{\idn{m}\idn{n}\idn{q}}} {\textcolor{olive}{X}{\idn{m}}} {\textcolor{olive}{p}{\idn{n}}} $}, so that 
{$ {\textcolor{olive}{Z}{\idn{k}}} $} gets expressed entirely in terms of 
{$ {\textcolor{olive}{p}{\idn{k}}} $}, 
{$ \textcolor{olive}{X} $} and 
{$ \textcolor{olive}{V} $}
\end{Maple Normal}
\mapleinput
{$ \displaystyle \texttt{>\,} \mathit{Simplify} (\mathit{SubstituteTensor} (\mapleref{(10)},\mapleref{(34)})) $}

\begin{dmath}\label{(35)}
{\textcolor{olive}{Z}{\idn{k}}}=\frac{-\mathrm{i} \hslash  {\textcolor{olive}{p}{\idn{k}}}+\kappa  \textcolor{olive}{V} {\textcolor{olive}{X}{\idn{k}}} m_{e}-{\textcolor{olive}{X}{\idn{k}}} {\textcolor{olive}{p}{\idn{m}\iup{2}}}+{\textcolor{olive}{X}{\idn{m}}} {\textcolor{olive}{p}{\idn{k}}} {\textcolor{olive}{p}{\idn{m}}}}{m_{e}}
\end{dmath}
\begin{Maple Normal}
Taking the commutator between \mapleref{(5)}
{$ \esequiv \textcolor{olive}{H}=\frac{ {\textcolor{olive}{p}{\idn{l}}}^{2}}{2m_{e}}-\kappa  \textcolor{olive}{V} $}, and this expression for 
{$ {\textcolor{olive}{Z}{\idn{k}}} $} we have the starting point towards showing that 
{$ [\textcolor{olive}{H},{\textcolor{olive}{Z}{\idn{k}}}]_{-}=0 $}
\end{Maple Normal}
\mapleinput
{$ \displaystyle \texttt{>\,}  \frac{2\,m_{e}}{\kappa  \hslash}\,\mathit{Simplify} (\mathit{Commutator} (\mapleref{(5)},\mapleref{(35)})) $}

\begin{dmath}\label{(36)}
\frac{2 m_{e} \left[\textcolor{olive}{H},{\textcolor{olive}{Z}{\idn{k}}}\right]_{-}}{\kappa  \hslash}=\hslash  {\textcolor{olive}{V}}^{3} {\textcolor{olive}{X}{\idn{k}}}+\hslash  {\textcolor{olive}{V}}^{5} {\textcolor{olive}{X}{\idn{a}\iup{2}}} {\textcolor{olive}{X}{\idn{k}}}+2 \mathrm{i} {\textcolor{olive}{X}{\idn{a}\iup{2}}} {\textcolor{olive}{p}{\idn{k}}} {\textcolor{olive}{V}}^{3}+2 \mathrm{i} \textcolor{olive}{V} {\textcolor{olive}{X}{\idn{a}}} {\textcolor{olive}{X}{\idn{k}}} {\textcolor{olive}{p}{\idn{a}}} {\textcolor{olive}{V}}^{2}-2 \mathrm{i} {\textcolor{olive}{X}{\idn{a}}} {\textcolor{olive}{X}{\idn{k}}} {\textcolor{olive}{p}{\idn{a}}} {\textcolor{olive}{V}}^{3}-2 \mathrm{i} {\textcolor{olive}{p}{\idn{k}}} \textcolor{olive}{V}
\end{dmath}
\begin{Maple Normal}
In order to use the identities
\end{Maple Normal}
\mapleinput
{$ \displaystyle \texttt{>\,} \mapleref{(9)},V (X)^{2}\cdot \,\mapleref{(9)} $}

\begin{dmath}\label{(37)}
{\textcolor{olive}{V}}^{3} {\textcolor{olive}{X}{\idn{l}\iup{2}}} \hiderel{=} \textcolor{olive}{V},{\textcolor{olive}{V}}^{5} {\textcolor{olive}{X}{\idn{l}\iup{2}}} \hiderel{=} {\textcolor{olive}{V}}^{3}
\end{dmath}
\begin{Maple Normal}
we sort the products using the ordering shown in the left-hand sides
\end{Maple Normal}
\mapleinput
{$ \displaystyle \texttt{>\,} \mathit{SortProducts} (\mapleref{(36)},[V (X)^{5},V (X)^{3},X [a]^{2}]) $}

\begin{dmath}\label{(38)}
\frac{2 m_{e} \left[\textcolor{olive}{H},{\textcolor{olive}{Z}{\idn{k}}}\right]_{-}}{\kappa  \hslash}=\hslash  {\textcolor{olive}{V}}^{3} {\textcolor{olive}{X}{\idn{k}}}-5 \hslash  {\textcolor{olive}{V}}^{5} {\textcolor{olive}{X}{\idn{a}\iup{2}}} {\textcolor{olive}{X}{\idn{k}}}+2 \mathrm{i} {\textcolor{olive}{V}}^{3} {\textcolor{olive}{X}{\idn{a}\iup{2}}} {\textcolor{olive}{p}{\idn{k}}}+2 \mathrm{i} \textcolor{olive}{V} {\textcolor{olive}{X}{\idn{a}}} {\textcolor{olive}{X}{\idn{k}}} {\textcolor{olive}{p}{\idn{a}}} {\textcolor{olive}{V}}^{2}-2 \mathrm{i} {\textcolor{olive}{X}{\idn{a}}} {\textcolor{olive}{X}{\idn{k}}} {\textcolor{olive}{p}{\idn{a}}} {\textcolor{olive}{V}}^{3}-2 \mathrm{i} {\textcolor{olive}{p}{\idn{k}}} \textcolor{olive}{V}
\end{dmath}
\mapleinput
{$ \displaystyle \texttt{>\,} \mathit{SubstituteTensor} (\mapleref{(37)},\mapleref{(38)}) $}

\begin{dmath}\label{(39)}
\frac{2 m_{e} \left[\textcolor{olive}{H},{\textcolor{olive}{Z}{\idn{k}}}\right]_{-}}{\kappa  \hslash}=-4 \hslash  {\textcolor{olive}{V}}^{3} {\textcolor{olive}{X}{\idn{k}}}+2 \mathrm{i} \textcolor{olive}{V} {\textcolor{olive}{p}{\idn{k}}}+2 \mathrm{i} \textcolor{olive}{V} {\textcolor{olive}{X}{\idn{a}}} {\textcolor{olive}{X}{\idn{k}}} {\textcolor{olive}{p}{\idn{a}}} {\textcolor{olive}{V}}^{2}-2 \mathrm{i} {\textcolor{olive}{X}{\idn{a}}} {\textcolor{olive}{X}{\idn{k}}} {\textcolor{olive}{p}{\idn{a}}} {\textcolor{olive}{V}}^{3}-2 \mathrm{i} {\textcolor{olive}{p}{\idn{k}}} \textcolor{olive}{V}
\end{dmath}
\mapleinput
{$ \displaystyle \texttt{>\,} \mathit{Simplify} (\mapleref{(39)}) $}

\begin{dmath}\label{(40)}
\frac{2 m_{e} \left[\textcolor{olive}{H},{\textcolor{olive}{Z}{\idn{k}}}\right]_{-}}{\kappa  \hslash}=-2 \hslash  \left({\textcolor{olive}{V}}^{3} {\textcolor{olive}{X}{\idn{k}}}-{\textcolor{olive}{V}}^{5} {\textcolor{olive}{X}{\idn{a}\iup{2}}} {\textcolor{olive}{X}{\idn{k}}}\right)
\end{dmath}
\mapleinput
{$ \displaystyle \texttt{>\,} \frac{\kappa  \hslash}{2 \mathit{m_{e}}}\,\mathit{SubstituteTensor} (\mapleref{(37)},\mapleref{(40)}) $}

\begin{dmath}\label{(41)}
\left[\textcolor{olive}{H},{\textcolor{olive}{Z}{\idn{k}}}\right]_{-}=0
\end{dmath}
\begin{Maple Normal}
And this is the result we wanted to prove.
\end{Maple Normal}
\subsection{\textbf{The commutator between the Hamiltonian and Runge-Lenz tensor: alternative derivation using differential operators}}
\begin{Maple Normal}
As done in the previous section when deriving the commutators between linear and angular momentum, on the one hand, and the central potential 
{$ \textcolor{olive}{V} $} on the other hand, the idea here is again to use differential operators taking advantage of the ability to compute with them as operands of a product, that get applied only when it appears convenient for us
\end{Maple Normal}
\mapleinput
{$ \displaystyle \texttt{>\,} \mathit{Setup} (\mathit{differentialoperators} =\{[p [k],[x ,y ,z]]\}) $}

\begin{dmath}\label{(42)}
\left[\mathit{differentialoperators} =\left\{\left[{\textcolor{olive}{p}{\idn{k}}},\left[\textcolor{olive}{X}\right]\right]\right\}\right]
\end{dmath}
\begin{Maple Normal}
So take the starting point \mapleref{(36)} 
\end{Maple Normal}
\mapleinput
{$ \displaystyle \texttt{>\,} \,\mapleref{(36)} $}

\begin{dmath}\label{(43)}
\frac{2 m_{e} \left[\textcolor{olive}{H},{\textcolor{olive}{Z}{\idn{k}}}\right]_{-}}{\kappa  \hslash}=\hslash  {\textcolor{olive}{V}}^{3} {\textcolor{olive}{X}{\idn{k}}}+\hslash  {\textcolor{olive}{V}}^{5} {\textcolor{olive}{X}{\idn{a}\iup{2}}} {\textcolor{olive}{X}{\idn{k}}}+2 \mathrm{i} {\textcolor{olive}{X}{\idn{a}\iup{2}}} {\textcolor{olive}{p}{\idn{k}}} {\textcolor{olive}{V}}^{3}+2 \mathrm{i} \textcolor{olive}{V} {\textcolor{olive}{X}{\idn{a}}} {\textcolor{olive}{X}{\idn{k}}} {\textcolor{olive}{p}{\idn{a}}} {\textcolor{olive}{V}}^{2}-2 \mathrm{i} {\textcolor{olive}{X}{\idn{a}}} {\textcolor{olive}{X}{\idn{k}}} {\textcolor{olive}{p}{\idn{a}}} {\textcolor{olive}{V}}^{3}-2 \mathrm{i} {\textcolor{olive}{p}{\idn{k}}} \textcolor{olive}{V}
\end{dmath}
\begin{Maple Normal}
and to show that the right-hand side is equal to 0, multiply by a generic function 
{$ G (\textcolor{olive}{X}) $} followed by transforming the products involving 
{$ \textcolor{olive}{p}_{n} $} into the application of this differential operator 
{$ \textcolor{olive}{p}_{n}=-i \hslash {\partial{\idn{n}}} $} 
\end{Maple Normal}
\mapleinput
{$ \displaystyle \texttt{>\,} \mapleref{(36)}\cdot G (X) $}

\begin{dmath}\label{(44)}
\frac{2 m_{e} \left[\textcolor{olive}{H},{\textcolor{olive}{Z}{\idn{k}}}\right]_{-} G}{\kappa  \hslash}=\left(\hslash  {\textcolor{olive}{V}}^{3} {\textcolor{olive}{X}{\idn{k}}}+\hslash  {\textcolor{olive}{V}}^{5} {\textcolor{olive}{X}{\idn{a}\iup{2}}} {\textcolor{olive}{X}{\idn{k}}}+2 \mathrm{i} {\textcolor{olive}{X}{\idn{a}\iup{2}}} {\textcolor{olive}{p}{\idn{k}}} {\textcolor{olive}{V}}^{3}+2 \mathrm{i} \textcolor{olive}{V} {\textcolor{olive}{X}{\idn{a}}} {\textcolor{olive}{X}{\idn{k}}} {\textcolor{olive}{p}{\idn{a}}} {\textcolor{olive}{V}}^{2}-2 \mathrm{i} {\textcolor{olive}{X}{\idn{a}}} {\textcolor{olive}{X}{\idn{k}}} {\textcolor{olive}{p}{\idn{a}}} {\textcolor{olive}{V}}^{3}-2 \mathrm{i} {\textcolor{olive}{p}{\idn{k}}} \textcolor{olive}{V}\right) G 
\end{dmath}
\mapleinput
{$ \displaystyle \texttt{>\,} \mathit{ApplyProductsOfDifferentialOperators} (\mapleref{(44)}) $}

\begin{dmath}\label{(45)}
\frac{2 m_{e} \left[\textcolor{olive}{H},{\textcolor{olive}{Z}{\idn{k}}}\right]_{-} G}{\kappa  \hslash}=-2 \hslash  {\textcolor{olive}{X}{\idn{a}}} {\textcolor{olive}{X}{\idn{k}}} \left(\left({\partial{\idn{a}}}\! \left(\textcolor{olive}{V}\right) {\textcolor{olive}{V}}^{2}+\textcolor{olive}{V} {\partial{\idn{a}}}\! \left(\textcolor{olive}{V}\right) \textcolor{olive}{V}+{\textcolor{olive}{V}}^{2} {\partial{\idn{a}}}\! \left(\textcolor{olive}{V}\right)\right) G +{\textcolor{olive}{V}}^{3} {\partial{\idn{a}}}\! \left(G \right)\right)-2 \hslash  \textcolor{olive}{V} {\partial{\idn{k}}}\! \left(G \right)-2 \hslash  {\partial{\idn{k}}}\! \left(\textcolor{olive}{V}\right) G +2 \hslash  {\textcolor{olive}{X}{\idn{a}\iup{2}}} \left(\left({\partial{\idn{k}}}\! \left(\textcolor{olive}{V}\right) {\textcolor{olive}{V}}^{2}+\textcolor{olive}{V} {\partial{\idn{k}}}\! \left(\textcolor{olive}{V}\right) \textcolor{olive}{V}+{\textcolor{olive}{V}}^{2} {\partial{\idn{k}}}\! \left(\textcolor{olive}{V}\right)\right) G +{\textcolor{olive}{V}}^{3} {\partial{\idn{k}}}\! \left(G \right)\right)+2 \hslash  \textcolor{olive}{V} {\textcolor{olive}{X}{\idn{a}}} {\textcolor{olive}{X}{\idn{k}}} \left(\left({\partial{\idn{a}}}\! \left(\textcolor{olive}{V}\right) \textcolor{olive}{V}+\textcolor{olive}{V} {\partial{\idn{a}}}\! \left(\textcolor{olive}{V}\right)\right) G +{\textcolor{olive}{V}}^{2} {\partial{\idn{a}}}\! \left(G \right)\right)+\hslash  {\textcolor{olive}{V}}^{3} {\textcolor{olive}{X}{\idn{k}}} G +\hslash  {\textcolor{olive}{V}}^{5} {\textcolor{olive}{X}{\idn{a}\iup{2}}} {\textcolor{olive}{X}{\idn{k}}} G 
\end{dmath}
\mapleinput
{$ \displaystyle \texttt{>\,} \frac{1}{\hslash}\,\mathit{Simplify} (\mapleref{(45)}) $}

\begin{dmath}\label{(46)}
\frac{2 m_{e} \left[\textcolor{olive}{H},{\textcolor{olive}{Z}{\idn{k}}}\right]_{-} G}{\hslash^{2} \kappa}=2 \textcolor{olive}{V} {\textcolor{olive}{X}{\idn{a}\iup{2}}} {\partial{\idn{k}}}\! \left(\textcolor{olive}{V}\right) G \textcolor{olive}{V}+2 {\textcolor{olive}{V}}^{2} {\textcolor{olive}{X}{\idn{a}\iup{2}}} {\partial{\idn{k}}}\! \left(\textcolor{olive}{V}\right) G +G {\textcolor{olive}{V}}^{5} {\textcolor{olive}{X}{\idn{a}\iup{2}}} {\textcolor{olive}{X}{\idn{k}}}+2 {\textcolor{olive}{V}}^{3} {\textcolor{olive}{X}{\idn{a}\iup{2}}} {\partial{\idn{k}}}\! \left(G \right)-2 \textcolor{olive}{V} {\partial{\idn{k}}}\! \left(G \right)-2 {\partial{\idn{k}}}\! \left(\textcolor{olive}{V}\right) G +G {\textcolor{olive}{V}}^{3} {\textcolor{olive}{X}{\idn{k}}}+2 {\textcolor{olive}{X}{\idn{a}\iup{2}}} {\partial{\idn{k}}}\! \left(\textcolor{olive}{V}\right) G {\textcolor{olive}{V}}^{2}-2 {\textcolor{olive}{X}{\idn{a}}} {\textcolor{olive}{X}{\idn{k}}} {\partial{\idn{a}}}\! \left(\textcolor{olive}{V}\right) G {\textcolor{olive}{V}}^{2}
\end{dmath}
\begin{Maple Normal}
In addition, consider the application of 
{$ {\textcolor{olive}{p}{\idn{l}}} $} to the test function 
{$ G (\textcolor{olive}{X}) $}
\end{Maple Normal}
\mapleinput
{$ \displaystyle \texttt{>\,} p [l]\,G (X) $}

\begin{dmath}\label{(47)}
{\textcolor{olive}{p}{\idn{l}}} G 
\end{dmath}
\mapleinput
{$ \displaystyle \texttt{>\,} \mapleref{(47)}=\,\mathit{ApplyProductsOfDifferentialOperators} (\mapleref{(47)}) $}

\begin{dmath}\label{(48)}
{\textcolor{olive}{p}{\idn{l}}} G =\mathrm{-i} \hslash  {\partial{\idn{l}}}\! \left(G \right)
\end{dmath}
\mapleinput
{$ \displaystyle \texttt{>\,} \mathit{isolate} (\mapleref{(48)},{\partial{\idn{l}}}(G (X))) $}

\begin{dmath}\label{(49)}
{\partial{\idn{l}}}\! \left(G \right)=\frac{\mathrm{i} {\textcolor{olive}{p}{\idn{l}}} G}{\hslash}
\end{dmath}
\begin{Maple Normal}
Using this identity \mapleref{(49)} together with the derived identity \mapleref{(8)}, followed by multiplying by 
{$ G (\textcolor{olive}{X})^{\mathit{-}1} $} to remove the test function from the equation, we get
\end{Maple Normal}
\mapleinput
{$ \displaystyle \texttt{>\,} \mathit{Simplify} (\mathit{SubstituteTensor} (\mapleref{(8)},\mapleref{(49)},\mapleref{(46)})\cdot G (X)^{\mathit{-}1}) $}

\begin{dmath}\label{(50)}
\frac{2 m_{e} \left[\textcolor{olive}{H},{\textcolor{olive}{Z}{\idn{k}}}\right]_{-}}{\hslash^{2} \kappa}=-3 {\textcolor{olive}{V}}^{5} {\textcolor{olive}{X}{\idn{a}\iup{2}}} {\textcolor{olive}{X}{\idn{k}}}+\frac{2 \mathrm{i} {\textcolor{olive}{V}}^{3} {\textcolor{olive}{X}{\idn{a}\iup{2}}} {\textcolor{olive}{p}{\idn{k}}}}{\hslash}-\frac{2 \mathrm{i} \textcolor{olive}{V} {\textcolor{olive}{p}{\idn{k}}}}{\hslash}+3 {\textcolor{olive}{V}}^{3} {\textcolor{olive}{X}{\idn{k}}}
\end{dmath}
\begin{Maple Normal}
Applying \mapleref{(37)}
{$ \esequiv {\textcolor{olive}{V}}^{3} {\textcolor{olive}{X}{\idn{l}\iup{2}}}=\textcolor{olive}{V},{\textcolor{olive}{V}}^{5} {\textcolor{olive}{X}{\idn{l}\iup{2}}}={\textcolor{olive}{V}}^{3} $}
\end{Maple Normal}
\mapleinput
{$ \displaystyle \texttt{>\,} \frac{\hslash^{2} \kappa}{2 \mathit{m_{e}}}\,\mathit{SubstituteTensor} (\mapleref{(37)},\mapleref{(50)}) $}

\begin{dmath}\label{(51)}
\left[\textcolor{olive}{H},{\textcolor{olive}{Z}{\idn{k}}}\right]_{-}=0
\end{dmath}
\begin{Maple Normal}
Add to the setup these derived commutation rules between the Hamiltonian, angular momentum and Runge-Lenz tensors
\end{Maple Normal}
\mapleinput
{$ \displaystyle \texttt{>\,} \mapleref{(29)},\mapleref{(51)} $}

\begin{dmath}\label{(52)}
\left[\textcolor{olive}{H},{\textcolor{olive}{L}{\idn{q}}}\right]_{-} \hiderel{=} 0,\left[\textcolor{olive}{H},{\textcolor{olive}{Z}{\idn{k}}}\right]_{-} \hiderel{=} 0
\end{dmath}
\mapleinput
{$ \displaystyle \texttt{>\,} \mathit{Setup} (\mapleref{(52)}) $}

\begin{dmath}\label{(53)}
\left[\mathit{algebrarules} =\left\{\left[\textcolor{olive}{H},{\textcolor{olive}{L}{\idn{q}}}\right]_{-} \hiderel{=} 0,\left[\textcolor{olive}{H},{\textcolor{olive}{Z}{\idn{k}}}\right]_{-} \hiderel{=} 0,\left[{\textcolor{olive}{L}{\idn{j}}},{\textcolor{olive}{L}{\idn{k}}}\right]_{-} \hiderel{=} \mathrm{i} \hslash  {\epsilon{\idn{j}\idn{k}\idn{n}}} {\textcolor{olive}{L}{\idn{n}}},\left[{\textcolor{olive}{L}{\idn{q}}},\textcolor{olive}{V}\right]_{-} \hiderel{=} 0,
\\
\left[{\textcolor{olive}{p}{\idn{j}}},{\textcolor{olive}{L}{\idn{k}}}\right]_{-}\hiderel{=}\mathrm{i} \hslash  {\epsilon{\idn{j}\idn{k}\idn{n}}} {\textcolor{olive}{p}{\idn{n}}},\left[{\textcolor{olive}{p}{\idn{k}}},{\textcolor{olive}{p}{\idn{l}}}\right]_{-} \hiderel{=} 0,\left[{\textcolor{olive}{p}{\idn{q}}},\textcolor{olive}{V}\right]_{-} \hiderel{=} \mathrm{i} \hslash  {\textcolor{olive}{V}}^{3} {\textcolor{olive}{X}{\idn{q}}},
\\
\left[{\textcolor{olive}{p}{\idn{q}}},{\textcolor{olive}{V}}^{3}\right]_{-}\hiderel{=}3 \mathrm{i} \hslash  {\textcolor{olive}{V}}^{5} {\textcolor{olive}{X}{\idn{q}}},\left[{\textcolor{olive}{X}{\idn{j}}},{\textcolor{olive}{L}{\idn{k}}}\right]_{-} \hiderel{=} \mathrm{i} \hslash  {\epsilon{\idn{j}\idn{k}\idn{n}}} {\textcolor{olive}{X}{\idn{n}}},\left[{\textcolor{olive}{X}{\idn{k}}},{\textcolor{olive}{p}{\idn{l}}}\right]_{-} \hiderel{=} \mathrm{i} \hslash  {g{\idn{k}\idn{l}}},
\\
\left[{\textcolor{olive}{X}{\idn{k}}},\textcolor{olive}{V}\right]_{-}\hiderel{=}0\right\}\right]
\end{dmath}
\begin{Maple Normal}
Reset \textit{differentialoperators} in order to proceed to the next section working without them
\end{Maple Normal}
\mapleinput
{$ \displaystyle \texttt{>\,} \mathit{Setup} (\mathit{differentialoperators} =\mathit{none}) $}

\begin{dmath}\label{(54)}
\left[\mathit{differentialoperators} =\mathit{none} \right]
\end{dmath}
\section{\textbf{Commutation rules between the angular momentum and the Runge-Lenz tensors}}
\begin{Maple Normal}
Departing from the definition of these tensors, introduced in the previous sections
\end{Maple Normal}
\mapleinput
{$ \displaystyle \texttt{>\,} \mapleref{(10)};\,\mapleref{(35)} $}

\begin{maplelatex}
\mapleinline{inert}{2d}{}
{\[{\textcolor{olive}{L}{\idn{q}}}={\epsilon{\idn{m}\idn{n}\idn{q}}} {\textcolor{olive}{X}{\idn{m}}} {\textcolor{olive}{p}{\idn{n}}}\]}
\end{maplelatex}
\begin{dmath}\label{(55)}
{\textcolor{olive}{Z}{\idn{k}}}=\frac{-\mathrm{i} \hslash  {\textcolor{olive}{p}{\idn{k}}}+\kappa  \textcolor{olive}{V} {\textcolor{olive}{X}{\idn{k}}} m_{e}-{\textcolor{olive}{X}{\idn{k}}} {\textcolor{olive}{p}{\idn{m}\iup{2}}}+{\textcolor{olive}{X}{\idn{m}}} {\textcolor{olive}{p}{\idn{k}}} {\textcolor{olive}{p}{\idn{m}}}}{m_{e}}
\end{dmath}
\begin{Maple Normal}
the left-hand side of the identity to be proved is the left-hand side of the commutator of these two equations
\end{Maple Normal}
\mapleinput
{$ \displaystyle \texttt{>\,} m_{e}\,\mathit{Commutator} (\mapleref{(10)},\mapleref{(35)}) $}

\begin{dmath}\label{(56)}
m_{e} \left[{\textcolor{olive}{L}{\idn{q}}},{\textcolor{olive}{Z}{\idn{k}}}\right]_{-}={\epsilon{\idn{m}\idn{n}\idn{q}}} \hslash  \left(\mathrm{i} {\textcolor{olive}{X}{\idn{m}}} \left(-{g{\idn{k}\idn{n}}} \textcolor{olive}{V}+{\textcolor{olive}{V}}^{3} {\textcolor{olive}{X}{\idn{n}}} {\textcolor{olive}{X}{\idn{k}}}\right) \kappa  m_{e}+\hslash  {g{\idn{k}\idn{m}}} {\textcolor{olive}{p}{\idn{n}}}-2 \mathrm{i} {\textcolor{olive}{X}{\idn{k}}} {\textcolor{olive}{p}{\idn{a}}} {\textcolor{olive}{p}{\idn{n}}} {g{\idn{a}\idn{m}}}-\mathrm{i} {\textcolor{olive}{X}{\idn{m}}} {\textcolor{olive}{p}{\idn{k}}} {\textcolor{olive}{p}{\idn{a}}} {g{\idn{a}\idn{n}}}+\mathrm{i} {\textcolor{olive}{X}{\idn{m}}} {\textcolor{olive}{p}{\idn{a}\iup{2}}} {g{\idn{k}\idn{n}}}+\mathrm{i} {\textcolor{olive}{X}{\idn{a}}} \left({g{\idn{a}\idn{m}}} {\textcolor{olive}{p}{\idn{k}}}+{g{\idn{k}\idn{m}}} {\textcolor{olive}{p}{\idn{a}}}\right) {\textcolor{olive}{p}{\idn{n}}}\right)
\end{dmath}
\mapleinput
{$ \displaystyle \texttt{>\,} \mathit{Simplify} (\mapleref{(56)}) $}

\begin{dmath}\label{(57)}
m_{e} \left[{\textcolor{olive}{L}{\idn{q}}},{\textcolor{olive}{Z}{\idn{k}}}\right]_{-}=-\hslash  \left(\mathrm{i} \textcolor{olive}{V} {\textcolor{olive}{X}{\idn{a}}} \kappa  m_{e}+\hslash  {\textcolor{olive}{p}{\idn{a}}}-\mathrm{i} {\textcolor{olive}{X}{\idn{a}}} {\textcolor{olive}{p}{\idn{m}\iup{2}}}+\mathrm{i} {\textcolor{olive}{X}{\idn{m}}} {\textcolor{olive}{p}{\idn{a}}} {\textcolor{olive}{p}{\idn{m}}}\right) {\epsilon{\idn{a}\idn{k}\idn{q}}}
\end{dmath}
\begin{Maple Normal}
By eye, the right-hand side of \mapleref{(57)} is similar to the right-hand side of the definition of 
{$ {\textcolor{olive}{Z}{\idn{k}}} $} in \mapleref{(55)}, so introduce this definition directly into the right-hand side of \mapleref{(57)}. For that purpose, isolate 
{$ {\textcolor{olive}{X}{\idn{k}}} {\textcolor{olive}{p}{\idn{m}\iup{2}}} $} 
\end{Maple Normal}
\mapleinput
{$ \displaystyle \texttt{>\,} \mathit{isolate} (\mapleref{(55)},X [k]\cdot p [m]^{2}) $}

\begin{dmath}\label{(58)}
{\textcolor{olive}{X}{\idn{k}}} {\textcolor{olive}{p}{\idn{m}\iup{2}}}=-{\textcolor{olive}{Z}{\idn{k}}} m_{e}-\mathrm{i} \hslash  {\textcolor{olive}{p}{\idn{k}}}+\kappa  \textcolor{olive}{V} {\textcolor{olive}{X}{\idn{k}}} m_{e}+{\textcolor{olive}{X}{\idn{m}}} {\textcolor{olive}{p}{\idn{k}}} {\textcolor{olive}{p}{\idn{m}}}
\end{dmath}
\mapleinput
{$ \displaystyle \texttt{>\,} \mathit{SubstituteTensor} (\mapleref{(58)},\mapleref{(57)}) $}

\begin{dmath}\label{(59)}
m_{e} \left[{\textcolor{olive}{L}{\idn{q}}},{\textcolor{olive}{Z}{\idn{k}}}\right]_{-}=\mathrm{i} \hslash  \left(-{\textcolor{olive}{Z}{\idn{a}}} m_{e}+{\textcolor{olive}{X}{\idn{b}}} {\textcolor{olive}{p}{\idn{a}}} {\textcolor{olive}{p}{\idn{b}}}-{\textcolor{olive}{X}{\idn{m}}} {\textcolor{olive}{p}{\idn{a}}} {\textcolor{olive}{p}{\idn{m}}}\right) {\epsilon{\idn{a}\idn{k}\idn{q}}}
\end{dmath}
\begin{Maple Normal}
Simplifying, we get the desired result, and we substitute the \textit{active} by the \textit{inert} form of \textit{Commutator} for posterior use of this formula without having the \textit{Commutator} automatically executed.
\end{Maple Normal}
\mapleinput
{$ \displaystyle \texttt{>\,} \mathit{Simplify} (\mapleref{(59)}) $}

\begin{dmath}\label{(60)}
m_{e} \left[{\textcolor{olive}{L}{\idn{q}}},{\textcolor{olive}{Z}{\idn{k}}}\right]_{-}=\mathrm{-i} \hslash  {\textcolor{olive}{Z}{\idn{a}}} m_{e} {\epsilon{\idn{a}\idn{k}\idn{q}}}
\end{dmath}
\mapleinput
{$ \displaystyle \texttt{>\,} \frac{1}{\mathit{m_{e}}}\,\mathit{subs} (\mathit{Commutator} =\,\mathit{\%\!Commutator} ,\mapleref{(60)}) $}

\begin{dmath}\label{(61)}
\left[{\textcolor{olive}{L}{\idn{q}}},{\textcolor{olive}{Z}{\idn{k}}}\right]_{-}=\mathrm{-i} \hslash  {\epsilon{\idn{a}\idn{k}\idn{q}}} {\textcolor{olive}{Z}{\idn{a}}}
\end{dmath}
\begin{Maple Normal}
Set now this algebra rule to be available to the system when convenient
\end{Maple Normal}
\mapleinput
{$ \displaystyle \texttt{>\,} \mathit{Setup} (\mapleref{(61)}) $}

\begin{dmath}\label{(62)}
\left[\mathit{algebrarules} =\left\{\left[\textcolor{olive}{H},{\textcolor{olive}{L}{\idn{q}}}\right]_{-} \hiderel{=} 0,\left[\textcolor{olive}{H},{\textcolor{olive}{Z}{\idn{k}}}\right]_{-} \hiderel{=} 0,\left[{\textcolor{olive}{L}{\idn{j}}},{\textcolor{olive}{L}{\idn{k}}}\right]_{-} \hiderel{=} \mathrm{i} \hslash  {\epsilon{\idn{j}\idn{k}\idn{n}}} {\textcolor{olive}{L}{\idn{n}}},
\\
\left[{\textcolor{olive}{L}{\idn{q}}},{\textcolor{olive}{Z}{\idn{k}}}\right]_{-}\hiderel{=}\mathrm{-i} \hslash  {\epsilon{\idn{a}\idn{k}\idn{q}}} {\textcolor{olive}{Z}{\idn{a}}},\left[{\textcolor{olive}{L}{\idn{q}}},\textcolor{olive}{V}\right]_{-} \hiderel{=} 0,\left[{\textcolor{olive}{p}{\idn{j}}},{\textcolor{olive}{L}{\idn{k}}}\right]_{-} \hiderel{=} \mathrm{i} \hslash  {\epsilon{\idn{j}\idn{k}\idn{n}}} {\textcolor{olive}{p}{\idn{n}}},\left[{\textcolor{olive}{p}{\idn{k}}},{\textcolor{olive}{p}{\idn{l}}}\right]_{-} \hiderel{=} 0,
\\
\left[{\textcolor{olive}{p}{\idn{q}}},\textcolor{olive}{V}\right]_{-}\hiderel{=}\mathrm{i} \hslash  {\textcolor{olive}{V}}^{3} {\textcolor{olive}{X}{\idn{q}}},\left[{\textcolor{olive}{p}{\idn{q}}},{\textcolor{olive}{V}}^{3}\right]_{-} \hiderel{=} 3 \mathrm{i} \hslash  {\textcolor{olive}{V}}^{5} {\textcolor{olive}{X}{\idn{q}}},\left[{\textcolor{olive}{X}{\idn{j}}},{\textcolor{olive}{L}{\idn{k}}}\right]_{-} \hiderel{=} \mathrm{i} \hslash  {\epsilon{\idn{j}\idn{k}\idn{n}}} {\textcolor{olive}{X}{\idn{n}}},
\\
\left[{\textcolor{olive}{X}{\idn{k}}},{\textcolor{olive}{p}{\idn{l}}}\right]_{-}\hiderel{=}\mathrm{i} \hslash  {g{\idn{k}\idn{l}}},\left[{\textcolor{olive}{X}{\idn{k}}},\textcolor{olive}{V}\right]_{-} \hiderel{=} 0\right\}\right]
\end{dmath}
\subsection{\textbf{The scalar product between the quantized angular momentum and the Runge-Lenz tensors}}
\begin{Maple Normal}
Classically, the orbital momentum is perpendicular to the elliptic plane of motion, while the Runge-Lenz vector lies in that plane, so that  
{$ {\moverset{\rightarrow}{L}}_{\mathit{Classical}}\cdot  $} 
{$ {\moverset{\rightarrow}{Z}}_{\mathit{Classical}}=0 $}. In quantum mechanics, from \mapleref{(61)}
{$ \esequiv {{\textcolor{gray}{[}\textcolor{olive}{L}_{\textcolor{olive}{q}},{\moverset{}{\textcolor{olive}{Z}}}_{k}\textcolor{gray}{]}}}_{-}\neq 0 $} but  
{$ {\moverset{\rightarrow}{\textcolor{olive}{L}}}\cdot  $} 
{$ {\moverset{\rightarrow}{\textcolor{olive}{Z}}} $} = 
{$ {\moverset{\rightarrow}{\textcolor{olive}{Z}}}\cdot  $} 
{$ {\moverset{\rightarrow}{\textcolor{olive}{L}}}=0 $} still holds. To verify that, take the definition \mapleref{(30)} of the quantum Runge-Lenz vector and multiply it by 
{$ {\textcolor{olive}{L}{\idn{k}}} $}
\end{Maple Normal}
\mapleinput
{$ \displaystyle \texttt{>\,} \mapleref{(30)}\,.\,L [k] $}

\begin{dmath}\label{(63)}
{\textcolor{olive}{L}{\idn{k}}} {\textcolor{olive}{Z}{\idn{k}}}=\frac{{\epsilon{\idn{a}\idn{b}\idn{k}}} \left({\textcolor{olive}{L}{\idn{a}}} {\textcolor{olive}{p}{\idn{b}}} {\textcolor{olive}{L}{\idn{k}}}-{\textcolor{olive}{p}{\idn{a}}} {\textcolor{olive}{L}{\idn{b}}} {\textcolor{olive}{L}{\idn{k}}}\right)}{2 m_{e}}+\kappa  \textcolor{olive}{V} {\textcolor{olive}{X}{\idn{k}}} {\textcolor{olive}{L}{\idn{k}}}
\end{dmath}
\mapleinput
{$ \displaystyle \texttt{>\,} \mathit{Simplify} (\mapleref{(63)}) $}

\begin{dmath}\label{(64)}
{\textcolor{olive}{L}{\idn{k}}} {\textcolor{olive}{Z}{\idn{k}}}=\kappa  {\textcolor{olive}{L}{\idn{a}}} \textcolor{olive}{V} {\textcolor{olive}{X}{\idn{a}}}
\end{dmath}
\begin{Maple Normal}
Using \mapleref{(10)}
{$ \esequiv  $}
{$ {\textcolor{olive}{L}{\idn{q}}}={\epsilon{\idn{m}\idn{n}\idn{q}}} {\textcolor{olive}{X}{\idn{m}}} {\textcolor{olive}{p}{\idn{n}}} $},
\end{Maple Normal}
\mapleinput
{$ \displaystyle \texttt{>\,} \mathit{lhs} (\mapleref{(64)})=\mathit{SubstituteTensor} (\mapleref{(10)},\mathit{rhs} (\mapleref{(64)})) $}

\begin{dmath}\label{(65)}
{\textcolor{olive}{L}{\idn{k}}} {\textcolor{olive}{Z}{\idn{k}}}=\kappa  {\epsilon{\idn{a}\idn{m}\idn{n}}} {\textcolor{olive}{X}{\idn{m}}} {\textcolor{olive}{p}{\idn{n}}} \textcolor{olive}{V} {\textcolor{olive}{X}{\idn{a}}}
\end{dmath}
\mapleinput
{$ \displaystyle \texttt{>\,} \mathit{Simplify} (\mapleref{(65)}) $}

\begin{dmath}\label{(66)}
{\textcolor{olive}{L}{\idn{k}}} {\textcolor{olive}{Z}{\idn{k}}}=0
\end{dmath}
\begin{Maple Normal}
and due to \mapleref{(61)}
{$ \esequiv [{\textcolor{olive}{L}{\idn{q}}},{\textcolor{olive}{Z}{\idn{k}}}]_{-}=-\hslash  \mathrm{i} {\textcolor{olive}{Z}{\idn{a}}} {\epsilon{\idn{a}\idn{k}\idn{q}}} $}, reversing the order in the product,
\end{Maple Normal}
\mapleinput
{$ \displaystyle \texttt{>\,} \mathit{SortProducts} (\mapleref{(66)},[Z [k],L [k]]) $}

\begin{dmath}\label{(67)}
{\textcolor{olive}{Z}{\idn{k}}} {\textcolor{olive}{L}{\idn{k}}}=0
\end{dmath}
\section{\textbf{Commutation rules between the components of the Runge-Lenz tensor}}
\begin{Maple Normal}
Here again the starting point is \mapleref{(35)}, the definition of the quantum Runge-Lenz tensor
\end{Maple Normal}
\mapleinput
{$ \displaystyle \texttt{>\,} \mathit{SubstituteTensorIndices} (k =\,q ,\mapleref{(35)}) $}

\begin{dmath}\label{(68)}
{\textcolor{olive}{Z}{\idn{q}}}=\frac{-\mathrm{i} \hslash  {\textcolor{olive}{p}{\idn{q}}}+\kappa  \textcolor{olive}{V} {\textcolor{olive}{X}{\idn{q}}} m_{e}+{\textcolor{olive}{X}{\idn{m}}} {\textcolor{olive}{p}{\idn{q}}} {\textcolor{olive}{p}{\idn{m}}}-{\textcolor{olive}{X}{\idn{q}}} {\textcolor{olive}{p}{\idn{m}\iup{2}}}}{m_{e}}
\end{dmath}
\begin{Maple Normal}
The commutator 
{$ [{\textcolor{olive}{Z}{\idn{k}}},{\textcolor{olive}{Z}{\idn{q}}}]_{-} $}is computed via
\end{Maple Normal}
\mapleinput
{$ \displaystyle \texttt{>\,} m_{e}^{2}\,(\mathit{lhs} =\mathit{Expand} @ \mathit{rhs})(\mathit{Commutator} (\mapleref{(35)},\mapleref{(68)})) $}

\begin{dmath}\label{(69)}
m_{e}^{2} \left[{\textcolor{olive}{Z}{\idn{k}}},{\textcolor{olive}{Z}{\idn{q}}}\right]_{-}=\hslash  \left(-\left({g{\idn{a}\idn{k}}} {\textcolor{olive}{p}{\idn{a}}} {\textcolor{olive}{p}{\idn{q}}}-{g{\idn{m}\idn{q}}} {\textcolor{olive}{p}{\idn{k}}} {\textcolor{olive}{p}{\idn{m}}}-{g{\idn{k}\idn{q}}} \left({\textcolor{olive}{p}{\idn{a}\iup{2}}}-{\textcolor{olive}{p}{\idn{m}\iup{2}}}\right)\right) \hslash +\mathrm{i} {g{\idn{m}\idn{q}}} {\textcolor{olive}{X}{\idn{m}}} {\textcolor{olive}{p}{\idn{a}\iup{2}}} {\textcolor{olive}{p}{\idn{k}}}+\mathrm{i} {g{\idn{a}\idn{m}}} {\textcolor{olive}{X}{\idn{a}}} {\textcolor{olive}{p}{\idn{k}}} {\textcolor{olive}{p}{\idn{m}}} {\textcolor{olive}{p}{\idn{q}}}+\mathrm{i} {g{\idn{k}\idn{q}}} {\textcolor{olive}{X}{\idn{m}}} {\textcolor{olive}{p}{\idn{a}\iup{2}}} {\textcolor{olive}{p}{\idn{m}}}+\mathrm{i} {g{\idn{m}\idn{q}}} {\textcolor{olive}{X}{\idn{a}}} {\textcolor{olive}{p}{\idn{a}}} {\textcolor{olive}{p}{\idn{k}}} {\textcolor{olive}{p}{\idn{m}}}-\mathrm{i} {g{\idn{a}\idn{k}}} {\textcolor{olive}{X}{\idn{m}}} {\textcolor{olive}{p}{\idn{a}}} {\textcolor{olive}{p}{\idn{m}}} {\textcolor{olive}{p}{\idn{q}}}-\mathrm{i} {g{\idn{a}\idn{m}}} {\textcolor{olive}{X}{\idn{m}}} {\textcolor{olive}{p}{\idn{a}}} {\textcolor{olive}{p}{\idn{k}}} {\textcolor{olive}{p}{\idn{q}}}-2 \mathrm{i} {g{\idn{a}\idn{m}}} {\textcolor{olive}{X}{\idn{q}}} {\textcolor{olive}{p}{\idn{a}}} {\textcolor{olive}{p}{\idn{k}}} {\textcolor{olive}{p}{\idn{m}}}-2 \mathrm{i} {g{\idn{m}\idn{q}}} {\textcolor{olive}{X}{\idn{k}}} {\textcolor{olive}{p}{\idn{a}\iup{2}}} {\textcolor{olive}{p}{\idn{m}}}+2 \mathrm{i} {g{\idn{a}\idn{k}}} {\textcolor{olive}{X}{\idn{q}}} {\textcolor{olive}{p}{\idn{m}\iup{2}}} {\textcolor{olive}{p}{\idn{a}}}-\mathrm{i} {g{\idn{a}\idn{k}}} {\textcolor{olive}{X}{\idn{a}}} {\textcolor{olive}{p}{\idn{m}\iup{2}}} {\textcolor{olive}{p}{\idn{q}}}-\mathrm{i} {g{\idn{k}\idn{q}}} {\textcolor{olive}{X}{\idn{a}}} {\textcolor{olive}{p}{\idn{m}\iup{2}}} {\textcolor{olive}{p}{\idn{a}}}+2 \mathrm{i} {g{\idn{a}\idn{m}}} {\textcolor{olive}{X}{\idn{k}}} {\textcolor{olive}{p}{\idn{a}}} {\textcolor{olive}{p}{\idn{m}}} {\textcolor{olive}{p}{\idn{q}}}+\mathrm{i} {\textcolor{olive}{V}}^{3} {\textcolor{olive}{X}{\idn{k}}} {\textcolor{olive}{X}{\idn{m}}} {\textcolor{olive}{X}{\idn{q}}} {\textcolor{olive}{p}{\idn{m}}} \kappa  m_{e}+3 \hslash  {\textcolor{olive}{V}}^{5} {\textcolor{olive}{X}{\idn{a}\iup{2}}} {\textcolor{olive}{X}{\idn{k}}} {\textcolor{olive}{X}{\idn{q}}} \kappa  m_{e}-3 \hslash  {\textcolor{olive}{V}}^{5} {\textcolor{olive}{X}{\idn{m}\iup{2}}} {\textcolor{olive}{X}{\idn{k}}} {\textcolor{olive}{X}{\idn{q}}} \kappa  m_{e}-2 \mathrm{i} {\textcolor{olive}{X}{\idn{k}}} {\textcolor{olive}{p}{\idn{m}}} {\textcolor{olive}{V}}^{3} {\textcolor{olive}{X}{\idn{m}}} {\textcolor{olive}{X}{\idn{q}}} \kappa  m_{e}-\mathrm{i} {\textcolor{olive}{X}{\idn{a}}} {\textcolor{olive}{p}{\idn{q}}} {\textcolor{olive}{V}}^{3} {\textcolor{olive}{X}{\idn{a}}} {\textcolor{olive}{X}{\idn{k}}} \kappa  m_{e}-\mathrm{i} {\textcolor{olive}{V}}^{3} {\textcolor{olive}{X}{\idn{a}}} {\textcolor{olive}{X}{\idn{q}}} {\textcolor{olive}{p}{\idn{a}}} {\textcolor{olive}{X}{\idn{k}}} \kappa  m_{e}+2 \mathrm{i} {\textcolor{olive}{X}{\idn{q}}} {\textcolor{olive}{p}{\idn{a}}} {\textcolor{olive}{V}}^{3} {\textcolor{olive}{X}{\idn{a}}} {\textcolor{olive}{X}{\idn{k}}} \kappa  m_{e}+\mathrm{i} {\textcolor{olive}{X}{\idn{m}}} {\textcolor{olive}{p}{\idn{k}}} {\textcolor{olive}{V}}^{3} {\textcolor{olive}{X}{\idn{m}}} {\textcolor{olive}{X}{\idn{q}}} \kappa  m_{e}+2 \mathrm{i} \textcolor{olive}{V} {\textcolor{olive}{X}{\idn{k}}} {\textcolor{olive}{p}{\idn{m}}} \kappa  m_{e} {g{\idn{m}\idn{q}}}-\mathrm{i} {\textcolor{olive}{X}{\idn{m}}} {\textcolor{olive}{p}{\idn{k}}} \textcolor{olive}{V} \kappa  m_{e} {g{\idn{m}\idn{q}}}-\mathrm{i} \textcolor{olive}{V} {\textcolor{olive}{X}{\idn{m}}} {\textcolor{olive}{p}{\idn{m}}} \kappa  m_{e} {g{\idn{k}\idn{q}}}-2 \mathrm{i} \textcolor{olive}{V} {\textcolor{olive}{X}{\idn{q}}} {\textcolor{olive}{p}{\idn{a}}} \kappa  m_{e} {g{\idn{a}\idn{k}}}+\mathrm{i} \textcolor{olive}{V} {\textcolor{olive}{X}{\idn{a}}} {\textcolor{olive}{p}{\idn{q}}} \kappa  m_{e} {g{\idn{a}\idn{k}}}+\mathrm{i} \textcolor{olive}{V} {\textcolor{olive}{X}{\idn{a}}} {\textcolor{olive}{p}{\idn{a}}} \kappa  m_{e} {g{\idn{k}\idn{q}}}\right)
\end{dmath}
\mapleinput
{$ \displaystyle \texttt{>\,} \frac{\mathrm{i}}{\hslash}\,\mathit{Simplify} (\mapleref{(69)}) $}

\begin{dmath}\label{(70)}
\frac{\mathrm{i} m_{e}^{2} \left[{\textcolor{olive}{Z}{\idn{k}}},{\textcolor{olive}{Z}{\idn{q}}}\right]_{-}}{\hslash}=-{\textcolor{olive}{X}{\idn{a}\iup{2}}} {\textcolor{olive}{X}{\idn{q}}} {\textcolor{olive}{p}{\idn{k}}} {\textcolor{olive}{V}}^{3} \kappa  m_{e}+{\textcolor{olive}{X}{\idn{a}\iup{2}}} {\textcolor{olive}{X}{\idn{k}}} {\textcolor{olive}{p}{\idn{q}}} {\textcolor{olive}{V}}^{3} \kappa  m_{e}-3 {\textcolor{olive}{X}{\idn{k}}} {\textcolor{olive}{p}{\idn{q}}} \textcolor{olive}{V} \kappa  m_{e}+3 {\textcolor{olive}{X}{\idn{q}}} {\textcolor{olive}{p}{\idn{k}}} \textcolor{olive}{V} \kappa  m_{e}-{\textcolor{olive}{X}{\idn{m}}} {\textcolor{olive}{p}{\idn{a}\iup{2}}} {\textcolor{olive}{p}{\idn{m}}} {g{\idn{k}\idn{q}}}+{\textcolor{olive}{X}{\idn{a}}} {\textcolor{olive}{p}{\idn{m}\iup{2}}} {\textcolor{olive}{p}{\idn{a}}} {g{\idn{k}\idn{q}}}+{\textcolor{olive}{X}{\idn{k}}} {\textcolor{olive}{p}{\idn{a}\iup{2}}} {\textcolor{olive}{p}{\idn{q}}}-{\textcolor{olive}{X}{\idn{q}}} {\textcolor{olive}{p}{\idn{a}\iup{2}}} {\textcolor{olive}{p}{\idn{k}}}
\end{dmath}
\begin{Maple Normal}
In order to use \mapleref{(9)}
{$ \textcolor{olive}{\esequiv}{\textcolor{olive}{V}}^{3} {\textcolor{olive}{X}{\idn{l}}}^{2}=\textcolor{olive}{V} $}, sort the products in \mapleref{(70)} using the ordering 
{$ {\textcolor{olive}{V}}^{3} {\textcolor{olive}{X}{\idn{a}}}^{2} $} 
\end{Maple Normal}
\mapleinput
{$ \displaystyle \texttt{>\,} \mathit{Normal} (\mathit{SortProducts} (\mapleref{(70)},[V (X)^{3},X [a]^{2}])) $}

\begin{dmath}\label{(71)}
\frac{\mathrm{i} m_{e}^{2} \left[{\textcolor{olive}{Z}{\idn{k}}},{\textcolor{olive}{Z}{\idn{q}}}\right]_{-}}{\hslash}={\textcolor{olive}{V}}^{3} {\textcolor{olive}{X}{\idn{a}\iup{2}}} {\textcolor{olive}{X}{\idn{k}}} {\textcolor{olive}{p}{\idn{q}}} \kappa  m_{e}-{\textcolor{olive}{V}}^{3} {\textcolor{olive}{X}{\idn{a}\iup{2}}} {\textcolor{olive}{X}{\idn{q}}} {\textcolor{olive}{p}{\idn{k}}} \kappa  m_{e}-3 {\textcolor{olive}{X}{\idn{k}}} {\textcolor{olive}{p}{\idn{q}}} \textcolor{olive}{V} \kappa  m_{e}+3 {\textcolor{olive}{X}{\idn{q}}} {\textcolor{olive}{p}{\idn{k}}} \textcolor{olive}{V} \kappa  m_{e}-{\textcolor{olive}{X}{\idn{m}}} {\textcolor{olive}{p}{\idn{a}\iup{2}}} {\textcolor{olive}{p}{\idn{m}}} {g{\idn{k}\idn{q}}}+{\textcolor{olive}{X}{\idn{a}}} {\textcolor{olive}{p}{\idn{m}\iup{2}}} {\textcolor{olive}{p}{\idn{a}}} {g{\idn{k}\idn{q}}}+{\textcolor{olive}{X}{\idn{k}}} {\textcolor{olive}{p}{\idn{a}\iup{2}}} {\textcolor{olive}{p}{\idn{q}}}-{\textcolor{olive}{X}{\idn{q}}} {\textcolor{olive}{p}{\idn{a}\iup{2}}} {\textcolor{olive}{p}{\idn{k}}}
\end{dmath}
\mapleinput
{$ \displaystyle \texttt{>\,} \mathit{SubstituteTensor} (\mapleref{(9)},\mapleref{(71)}) $}

\begin{dmath}\label{(72)}
\frac{\mathrm{i} m_{e}^{2} \left[{\textcolor{olive}{Z}{\idn{k}}},{\textcolor{olive}{Z}{\idn{q}}}\right]_{-}}{\hslash}=\textcolor{olive}{V} {\textcolor{olive}{X}{\idn{k}}} {\textcolor{olive}{p}{\idn{q}}} \kappa  m_{e}-\kappa  m_{e} \textcolor{olive}{V} {\textcolor{olive}{X}{\idn{q}}} {\textcolor{olive}{p}{\idn{k}}}-3 {\textcolor{olive}{X}{\idn{k}}} {\textcolor{olive}{p}{\idn{q}}} \textcolor{olive}{V} \kappa  m_{e}+3 {\textcolor{olive}{X}{\idn{q}}} {\textcolor{olive}{p}{\idn{k}}} \textcolor{olive}{V} \kappa  m_{e}-{\textcolor{olive}{X}{\idn{m}}} {\textcolor{olive}{p}{\idn{a}\iup{2}}} {\textcolor{olive}{p}{\idn{m}}} {g{\idn{k}\idn{q}}}+{\textcolor{olive}{X}{\idn{a}}} {\textcolor{olive}{p}{\idn{m}\iup{2}}} {\textcolor{olive}{p}{\idn{a}}} {g{\idn{k}\idn{q}}}+{\textcolor{olive}{X}{\idn{k}}} {\textcolor{olive}{p}{\idn{a}\iup{2}}} {\textcolor{olive}{p}{\idn{q}}}-{\textcolor{olive}{X}{\idn{q}}} {\textcolor{olive}{p}{\idn{a}\iup{2}}} {\textcolor{olive}{p}{\idn{k}}}
\end{dmath}
\begin{Maple Normal}
Regarding the term quadratic in the momentum, from the expression for the Hamiltonian \mapleref{(5)}
{$ \esequiv \textcolor{olive}{H}=\frac{ {\textcolor{olive}{p}{\idn{l}}}^{2}}{2m_{e}}-\kappa  \textcolor{olive}{V} $},
\end{Maple Normal}
\mapleinput
{$ \displaystyle \texttt{>\,} \mathit{isolate} (\mapleref{(5)},p [l]^{2}) $}

\begin{dmath}\label{(73)}
{\textcolor{olive}{p}{\idn{l}\iup{2}}}=2 \left(\kappa  \textcolor{olive}{V}+\textcolor{olive}{H}\right) m_{e}
\end{dmath}
\begin{Maple Normal}
In order to use this equation \mapleref{(73)} to substitute 
{$ {\textcolor{olive}{p}{\idn{l}\iup{2}}} $} into the expression \mapleref{(72)} for 
{$ [{\textcolor{olive}{Z}{\idn{k}}},{\textcolor{olive}{Z}{\idn{q}}}]_{-} $} and \textit{not} receive noncommutative products with 
{$ \textcolor{olive}{H} $} in between the position 
{$ {\textcolor{olive}{X}{\idn{k}}} $} and momentum 
{$ {\textcolor{olive}{p}{\idn{q}}} $}operators (that would require further using, afterwards, of the commutator between 
{$ \textcolor{olive}{H} $}and 
{$ \textcolor{olive}{p}_{q} $}) , sort first the products in \mapleref{(72)} positioning all square of momentums  
{$ {\textcolor{olive}{p}{\idn{}\iup{2}}} $} to the right of occurrences of 
{$ {\textcolor{olive}{p}{\idn{}}} $} 
\end{Maple Normal}
\mapleinput
{$ \displaystyle \texttt{>\,} \mathit{SortProducts} (\mapleref{(72)},[p [a],p [k],p [m],p [q],p [a]^{2},p [m]^{2}]) $}

\begin{dmath}\label{(74)}
\frac{\mathrm{i} m_{e}^{2} \left[{\textcolor{olive}{Z}{\idn{k}}},{\textcolor{olive}{Z}{\idn{q}}}\right]_{-}}{\hslash}=\textcolor{olive}{V} {\textcolor{olive}{X}{\idn{k}}} {\textcolor{olive}{p}{\idn{q}}} \kappa  m_{e}-\kappa  m_{e} \textcolor{olive}{V} {\textcolor{olive}{X}{\idn{q}}} {\textcolor{olive}{p}{\idn{k}}}-3 {\textcolor{olive}{X}{\idn{k}}} {\textcolor{olive}{p}{\idn{q}}} \textcolor{olive}{V} \kappa  m_{e}+3 {\textcolor{olive}{X}{\idn{q}}} {\textcolor{olive}{p}{\idn{k}}} \textcolor{olive}{V} \kappa  m_{e}-{\textcolor{olive}{X}{\idn{m}}} {\textcolor{olive}{p}{\idn{m}}} {\textcolor{olive}{p}{\idn{a}\iup{2}}} {g{\idn{k}\idn{q}}}+{\textcolor{olive}{X}{\idn{a}}} {\textcolor{olive}{p}{\idn{a}}} {\textcolor{olive}{p}{\idn{m}\iup{2}}} {g{\idn{k}\idn{q}}}+{\textcolor{olive}{X}{\idn{k}}} {\textcolor{olive}{p}{\idn{q}}} {\textcolor{olive}{p}{\idn{a}\iup{2}}}-{\textcolor{olive}{X}{\idn{q}}} {\textcolor{olive}{p}{\idn{k}}} {\textcolor{olive}{p}{\idn{a}\iup{2}}}
\end{dmath}
\mapleinput
{$ \displaystyle \texttt{>\,} \mathit{SubstituteTensor} (\mapleref{(73)},\mapleref{(74)}) $}

\begin{dmath}\label{(75)}
\frac{\mathrm{i} m_{e}^{2} \left[{\textcolor{olive}{Z}{\idn{k}}},{\textcolor{olive}{Z}{\idn{q}}}\right]_{-}}{\hslash}=\textcolor{olive}{V} {\textcolor{olive}{X}{\idn{k}}} {\textcolor{olive}{p}{\idn{q}}} \kappa  m_{e}-\kappa  m_{e} \textcolor{olive}{V} {\textcolor{olive}{X}{\idn{q}}} {\textcolor{olive}{p}{\idn{k}}}-3 {\textcolor{olive}{X}{\idn{k}}} {\textcolor{olive}{p}{\idn{q}}} \textcolor{olive}{V} \kappa  m_{e}+3 {\textcolor{olive}{X}{\idn{q}}} {\textcolor{olive}{p}{\idn{k}}} \textcolor{olive}{V} \kappa  m_{e}-{\textcolor{olive}{X}{\idn{m}}} {\textcolor{olive}{p}{\idn{m}}} 2 \left(\kappa  \textcolor{olive}{V}+\textcolor{olive}{H}\right) m_{e} {g{\idn{k}\idn{q}}}+{\textcolor{olive}{X}{\idn{a}}} {\textcolor{olive}{p}{\idn{a}}} 2 \left(\kappa  \textcolor{olive}{V}+\textcolor{olive}{H}\right) m_{e} {g{\idn{k}\idn{q}}}+{\textcolor{olive}{X}{\idn{k}}} {\textcolor{olive}{p}{\idn{q}}} 2 \left(\kappa  \textcolor{olive}{V}+\textcolor{olive}{H}\right) m_{e}-{\textcolor{olive}{X}{\idn{q}}} {\textcolor{olive}{p}{\idn{k}}} 2 \left(\kappa  \textcolor{olive}{V}+\textcolor{olive}{H}\right) m_{e}
\end{dmath}
\mapleinput
{$ \displaystyle \texttt{>\,} \frac{\hslash}{\mathit{\mathrm{i}}\,\mathit{m_{e}^{2}}}\,\mathit{Simplify} (\mapleref{(75)}) $}

\begin{dmath}\label{(76)}
\left[{\textcolor{olive}{Z}{\idn{k}}},{\textcolor{olive}{Z}{\idn{q}}}\right]_{-}=\frac{2 \mathrm{i} \hslash  \left(-{\textcolor{olive}{X}{\idn{k}}} {\textcolor{olive}{p}{\idn{q}}} \textcolor{olive}{H}+{\textcolor{olive}{X}{\idn{q}}} {\textcolor{olive}{p}{\idn{k}}} \textcolor{olive}{H}\right)}{m_{e}}
\end{dmath}
\begin{Maple Normal}
Finally, from the definition of the angular momentum \mapleref{(10)}
{$ \esequiv {\textcolor{olive}{L}{\idn{q}}}={\epsilon{\idn{m}\idn{n}\idn{q}}} {\textcolor{olive}{X}{\idn{m}}} {\textcolor{olive}{p}{\idn{n}}} $}, multiplying by 
{$ {\epsilon{\idn{a}\idn{b}\idn{c}}} $} we can construct an expression for 
{$ {\textcolor{olive}{X}{\idn{a}}} {\textcolor{olive}{p}{\idn{b}}} \textcolor{olive}{H}-{\textcolor{olive}{X}{\idn{b}}} {\textcolor{olive}{p}{\idn{a}}} \textcolor{olive}{H} $} in terms of 
{$ {\textcolor{olive}{L}{\idn{q}}} $}
\end{Maple Normal}
\mapleinput
{$ \displaystyle \texttt{>\,} \mathit{LeviCivita} [a ,b ,q]\cdot \mapleref{(10)} $}

\begin{dmath}\label{(77)}
{\epsilon{\idn{a}\idn{b}\idn{q}}} {\textcolor{olive}{L}{\idn{q}}}={\epsilon{\idn{a}\idn{b}\idn{q}}} {\epsilon{\idn{m}\idn{n}\idn{q}}} {\textcolor{olive}{X}{\idn{m}}} {\textcolor{olive}{p}{\idn{n}}}
\end{dmath}
\mapleinput
{$ \displaystyle \texttt{>\,} \mathit{Simplify} ((\mathit{rhs} =\mathit{lhs})(\mapleref{(77)})) $}

\begin{dmath}\label{(78)}
{\textcolor{olive}{X}{\idn{a}}} {\textcolor{olive}{p}{\idn{b}}}-{\textcolor{olive}{X}{\idn{b}}} {\textcolor{olive}{p}{\idn{a}}}={\epsilon{\idn{a}\idn{b}\idn{q}}} {\textcolor{olive}{L}{\idn{q}}}
\end{dmath}
\mapleinput
{$ \displaystyle \texttt{>\,} \mathit{Expand} (\mapleref{(78)}\cdot H) $}

\begin{dmath}\label{(79)}
{\textcolor{olive}{X}{\idn{a}}} {\textcolor{olive}{p}{\idn{b}}} \textcolor{olive}{H}-{\textcolor{olive}{X}{\idn{b}}} {\textcolor{olive}{p}{\idn{a}}} \textcolor{olive}{H}={\epsilon{\idn{a}\idn{b}\idn{q}}} \textcolor{olive}{H} {\textcolor{olive}{L}{\idn{q}}}
\end{dmath}
\mapleinput
{$ \displaystyle \texttt{>\,} \mathit{SubstituteTensor} (\mapleref{(79)},\mapleref{(76)}) $}

\begin{dmath}\label{(80)}
\left[{\textcolor{olive}{Z}{\idn{k}}},{\textcolor{olive}{Z}{\idn{q}}}\right]_{-}=\frac{-2 \mathrm{i} \hslash  {\epsilon{\idn{c}\idn{k}\idn{q}}} \textcolor{olive}{H} {\textcolor{olive}{L}{\idn{c}}}}{m_{e}}
\end{dmath}
\begin{Maple Normal}
Which is the identity we wanted to prove.
\end{Maple Normal}
\subsection{\textbf{Alternative derivation using differential operators}}
\begin{Maple Normal}
Set again the \textit{differentialoperator} representation for the momentum operator 
{$ {\textcolor{olive}{p}{\idn{k}}} $}
\end{Maple Normal}
\mapleinput
{$ \displaystyle \texttt{>\,} \mathit{Setup} (\mathit{differentialoperators} =\{[p [k],[x ,y ,z]]\}) $}

\begin{dmath}\label{(81)}
\left[\mathit{differentialoperators} =\left\{\left[{\textcolor{olive}{p}{\idn{k}}},\left[\textcolor{olive}{X}\right]\right]\right\}\right]
\end{dmath}
\begin{Maple Normal}
and apply the expression \mapleref{(69)} for 
{$ [{\textcolor{olive}{Z}{\idn{k}}},{\textcolor{olive}{Z}{\idn{q}}}]_{-} $}to the test function 
{$ G (\textcolor{olive}{X}) $}
\end{Maple Normal}
\mapleinput
{$ \displaystyle \texttt{>\,} \mathit{ApplyProductsOfDifferentialOperators} (\mapleref{(69)}\cdot G (X)) $}

\begin{dmath}\label{(82)}
m_{e}^{2} \left[{\textcolor{olive}{Z}{\idn{k}}},{\textcolor{olive}{Z}{\idn{q}}}\right]_{-} G =\hslash  \left(-\kappa  m_{e} \hslash  {\textcolor{olive}{X}{\idn{a}}} \left(\left({\partial{\idn{q}}}\! \left(\textcolor{olive}{V}\right) {\textcolor{olive}{V}}^{2}+\textcolor{olive}{V} {\partial{\idn{q}}}\! \left(\textcolor{olive}{V}\right) \textcolor{olive}{V}+{\textcolor{olive}{V}}^{2} {\partial{\idn{q}}}\! \left(\textcolor{olive}{V}\right)\right) {\textcolor{olive}{X}{\idn{a}}} {\textcolor{olive}{X}{\idn{k}}} G +{g{\idn{a}\idn{q}}} {\textcolor{olive}{V}}^{3} {\textcolor{olive}{X}{\idn{k}}} G +{g{\idn{k}\idn{q}}} {\textcolor{olive}{V}}^{3} {\textcolor{olive}{X}{\idn{a}}} G +{\textcolor{olive}{V}}^{3} {\textcolor{olive}{X}{\idn{a}}} {\textcolor{olive}{X}{\idn{k}}} {\partial{\idn{q}}}\! \left(G \right)\right)-\kappa  m_{e} \hslash  {\textcolor{olive}{V}}^{3} {\textcolor{olive}{X}{\idn{a}}} {\textcolor{olive}{X}{\idn{q}}} \left({g{\idn{a}\idn{k}}} G +{\textcolor{olive}{X}{\idn{k}}} {\partial{\idn{a}}}\! \left(G \right)\right)+2 \kappa  m_{e} \hslash  {\textcolor{olive}{X}{\idn{q}}} \left(\left({\partial{\idn{a}}}\! \left(\textcolor{olive}{V}\right) {\textcolor{olive}{V}}^{2}+\textcolor{olive}{V} {\partial{\idn{a}}}\! \left(\textcolor{olive}{V}\right) \textcolor{olive}{V}+{\textcolor{olive}{V}}^{2} {\partial{\idn{a}}}\! \left(\textcolor{olive}{V}\right)\right) {\textcolor{olive}{X}{\idn{a}}} {\textcolor{olive}{X}{\idn{k}}} G +4 {\textcolor{olive}{V}}^{3} {\textcolor{olive}{X}{\idn{k}}} G +{\textcolor{olive}{V}}^{3} {\textcolor{olive}{X}{\idn{a}}} {\textcolor{olive}{X}{\idn{k}}} {\partial{\idn{a}}}\! \left(G \right)\right)+\kappa  m_{e} \hslash  {\textcolor{olive}{X}{\idn{m}}} \left(\left({\partial{\idn{k}}}\! \left(\textcolor{olive}{V}\right) {\textcolor{olive}{V}}^{2}+\textcolor{olive}{V} {\partial{\idn{k}}}\! \left(\textcolor{olive}{V}\right) \textcolor{olive}{V}+{\textcolor{olive}{V}}^{2} {\partial{\idn{k}}}\! \left(\textcolor{olive}{V}\right)\right) {\textcolor{olive}{X}{\idn{m}}} {\textcolor{olive}{X}{\idn{q}}} G +{g{\idn{k}\idn{m}}} {\textcolor{olive}{V}}^{3} {\textcolor{olive}{X}{\idn{q}}} G +{g{\idn{k}\idn{q}}} {\textcolor{olive}{V}}^{3} {\textcolor{olive}{X}{\idn{m}}} G +{\textcolor{olive}{V}}^{3} {\textcolor{olive}{X}{\idn{m}}} {\textcolor{olive}{X}{\idn{q}}} {\partial{\idn{k}}}\! \left(G \right)\right)+\kappa  m_{e} \hslash  {\textcolor{olive}{V}}^{3} {\textcolor{olive}{X}{\idn{k}}} {\textcolor{olive}{X}{\idn{m}}} {\textcolor{olive}{X}{\idn{q}}} {\partial{\idn{m}}}\! \left(G \right)-2 \kappa  m_{e} \hslash  {\textcolor{olive}{X}{\idn{k}}} \left(\left({\partial{\idn{m}}}\! \left(\textcolor{olive}{V}\right) {\textcolor{olive}{V}}^{2}+\textcolor{olive}{V} {\partial{\idn{m}}}\! \left(\textcolor{olive}{V}\right) \textcolor{olive}{V}+{\textcolor{olive}{V}}^{2} {\partial{\idn{m}}}\! \left(\textcolor{olive}{V}\right)\right) {\textcolor{olive}{X}{\idn{m}}} {\textcolor{olive}{X}{\idn{q}}} G +4 {\textcolor{olive}{V}}^{3} {\textcolor{olive}{X}{\idn{q}}} G +{\textcolor{olive}{V}}^{3} {\textcolor{olive}{X}{\idn{m}}} {\textcolor{olive}{X}{\idn{q}}} {\partial{\idn{m}}}\! \left(G \right)\right)-{g{\idn{k}\idn{q}}} \hslash^{3} {\textcolor{olive}{X}{\idn{m}}} {\partial{\idn{m}}}\! \left(\square \! \left(G \right)\right)-\hslash^{3} {g{\idn{m}\idn{q}}} {\textcolor{olive}{X}{\idn{a}}} {\partial{\idn{a}}}\! \left({\partial{\idn{k}}}\! \left({\partial{\idn{m}}}\! \left(G \right)\right)\right)+{g{\idn{k}\idn{q}}} \hslash^{3} {\textcolor{olive}{X}{\idn{a}}} {\partial{\idn{a}}}\! \left(\square \! \left(G \right)\right)-2 \hslash^{3} {g{\idn{a}\idn{m}}} {\textcolor{olive}{X}{\idn{k}}} {\partial{\idn{a}}}\! \left({\partial{\idn{m}}}\! \left({\partial{\idn{q}}}\! \left(G \right)\right)\right)+2 \hslash^{3} {g{\idn{m}\idn{q}}} {\textcolor{olive}{X}{\idn{k}}} {\partial{\idn{m}}}\! \left(\square \! \left(G \right)\right)+\hslash^{3} {g{\idn{a}\idn{k}}} {\textcolor{olive}{X}{\idn{a}}} {\partial{\idn{q}}}\! \left(\square \! \left(G \right)\right)+\hslash^{3} {g{\idn{a}\idn{k}}} {\textcolor{olive}{X}{\idn{m}}} {\partial{\idn{a}}}\! \left({\partial{\idn{m}}}\! \left({\partial{\idn{q}}}\! \left(G \right)\right)\right)+\hslash^{3} {g{\idn{a}\idn{m}}} {\textcolor{olive}{X}{\idn{m}}} {\partial{\idn{a}}}\! \left({\partial{\idn{k}}}\! \left({\partial{\idn{q}}}\! \left(G \right)\right)\right)-2 \hslash^{3} {g{\idn{a}\idn{k}}} {\textcolor{olive}{X}{\idn{q}}} {\partial{\idn{a}}}\! \left(\square \! \left(G \right)\right)-\hslash^{3} {g{\idn{a}\idn{m}}} {\textcolor{olive}{X}{\idn{a}}} {\partial{\idn{k}}}\! \left({\partial{\idn{m}}}\! \left({\partial{\idn{q}}}\! \left(G \right)\right)\right)-{g{\idn{m}\idn{q}}} \hslash^{3} {\partial{\idn{k}}}\! \left({\partial{\idn{m}}}\! \left(G \right)\right)+{g{\idn{a}\idn{k}}} \hslash^{3} {\partial{\idn{a}}}\! \left({\partial{\idn{q}}}\! \left(G \right)\right)-\hslash^{3} {g{\idn{m}\idn{q}}} {\textcolor{olive}{X}{\idn{m}}} {\partial{\idn{k}}}\! \left(\square \! \left(G \right)\right)+2 \hslash^{3} {g{\idn{a}\idn{m}}} {\textcolor{olive}{X}{\idn{q}}} {\partial{\idn{a}}}\! \left({\partial{\idn{k}}}\! \left({\partial{\idn{m}}}\! \left(G \right)\right)\right)+-3 \hslash  {\textcolor{olive}{V}}^{5} {\textcolor{olive}{X}{\idn{m}\iup{2}}} {\textcolor{olive}{X}{\idn{k}}} {\textcolor{olive}{X}{\idn{q}}} \kappa  m_{e} G +3 \hslash  {\textcolor{olive}{V}}^{5} {\textcolor{olive}{X}{\idn{a}\iup{2}}} {\textcolor{olive}{X}{\idn{k}}} {\textcolor{olive}{X}{\idn{q}}} \kappa  m_{e} G +\kappa  m_{e} {g{\idn{a}\idn{k}}} \hslash  \textcolor{olive}{V} {\textcolor{olive}{X}{\idn{a}}} {\partial{\idn{q}}}\! \left(G \right)+\kappa  m_{e} {g{\idn{k}\idn{q}}} \hslash  \textcolor{olive}{V} {\textcolor{olive}{X}{\idn{a}}} {\partial{\idn{a}}}\! \left(G \right)+2 \kappa  m_{e} {g{\idn{m}\idn{q}}} \hslash  \textcolor{olive}{V} {\textcolor{olive}{X}{\idn{k}}} {\partial{\idn{m}}}\! \left(G \right)-2 \kappa  m_{e} {g{\idn{a}\idn{k}}} \hslash  \textcolor{olive}{V} {\textcolor{olive}{X}{\idn{q}}} {\partial{\idn{a}}}\! \left(G \right)-\kappa  m_{e} {g{\idn{m}\idn{q}}} \hslash  {\textcolor{olive}{X}{\idn{m}}} \left({\partial{\idn{k}}}\! \left(\textcolor{olive}{V}\right) G +\textcolor{olive}{V} {\partial{\idn{k}}}\! \left(G \right)\right)-\kappa  m_{e} {g{\idn{k}\idn{q}}} \hslash  \textcolor{olive}{V} {\textcolor{olive}{X}{\idn{m}}} {\partial{\idn{m}}}\! \left(G \right)\right)
\end{dmath}
\mapleinput
{$ \displaystyle \texttt{>\,} \frac{1}{3\,\hslash^{2}}\,\mathit{Simplify} (\mapleref{(82)}) $}

\begin{dmath}\label{(83)}
\frac{m_{e}^{2} \left[{\textcolor{olive}{Z}{\idn{k}}},{\textcolor{olive}{Z}{\idn{q}}}\right]_{-} G}{3 \hslash^{2}}=\frac{\kappa  m_{e} \textcolor{olive}{V} {\textcolor{olive}{X}{\idn{d}}} {\partial{\idn{k}}}\! \left(\textcolor{olive}{V}\right) G \textcolor{olive}{V} {\textcolor{olive}{X}{\idn{q}}} {\textcolor{olive}{X}{\idn{d}}}}{3}-\frac{\kappa  m_{e} \textcolor{olive}{V} {\textcolor{olive}{X}{\idn{d}}} {\partial{\idn{q}}}\! \left(\textcolor{olive}{V}\right) G \textcolor{olive}{V} {\textcolor{olive}{X}{\idn{k}}} {\textcolor{olive}{X}{\idn{d}}}}{3}-\frac{2 \kappa  m_{e} \textcolor{olive}{V} {\textcolor{olive}{X}{\idn{k}}} {\partial{\idn{d}}}\! \left(\textcolor{olive}{V}\right) G \textcolor{olive}{V} {\textcolor{olive}{X}{\idn{q}}} {\textcolor{olive}{X}{\idn{d}}}}{3}+\frac{2 \kappa  m_{e} \textcolor{olive}{V} {\textcolor{olive}{X}{\idn{q}}} {\partial{\idn{d}}}\! \left(\textcolor{olive}{V}\right) G \textcolor{olive}{V} {\textcolor{olive}{X}{\idn{k}}} {\textcolor{olive}{X}{\idn{d}}}}{3}+\frac{\kappa  m_{e} {\textcolor{olive}{V}}^{2} {\textcolor{olive}{X}{\idn{d}}} {\partial{\idn{k}}}\! \left(\textcolor{olive}{V}\right) G {\textcolor{olive}{X}{\idn{q}}} {\textcolor{olive}{X}{\idn{d}}}}{3}-\frac{\kappa  m_{e} {\textcolor{olive}{V}}^{2} {\textcolor{olive}{X}{\idn{d}}} {\partial{\idn{q}}}\! \left(\textcolor{olive}{V}\right) G {\textcolor{olive}{X}{\idn{k}}} {\textcolor{olive}{X}{\idn{d}}}}{3}-\frac{2 \kappa  m_{e} {\textcolor{olive}{V}}^{2} {\textcolor{olive}{X}{\idn{k}}} {\partial{\idn{d}}}\! \left(\textcolor{olive}{V}\right) G {\textcolor{olive}{X}{\idn{q}}} {\textcolor{olive}{X}{\idn{d}}}}{3}+\frac{2 \kappa  m_{e} {\textcolor{olive}{V}}^{2} {\textcolor{olive}{X}{\idn{q}}} {\partial{\idn{d}}}\! \left(\textcolor{olive}{V}\right) G {\textcolor{olive}{X}{\idn{k}}} {\textcolor{olive}{X}{\idn{d}}}}{3}+\frac{\kappa  m_{e} {\textcolor{olive}{X}{\idn{d}}} {\partial{\idn{k}}}\! \left(\textcolor{olive}{V}\right) G {\textcolor{olive}{V}}^{2} {\textcolor{olive}{X}{\idn{q}}} {\textcolor{olive}{X}{\idn{d}}}}{3}-\frac{\kappa  m_{e} {\textcolor{olive}{X}{\idn{d}}} {\partial{\idn{q}}}\! \left(\textcolor{olive}{V}\right) G {\textcolor{olive}{V}}^{2} {\textcolor{olive}{X}{\idn{k}}} {\textcolor{olive}{X}{\idn{d}}}}{3}-\frac{2 \kappa  m_{e} {\textcolor{olive}{X}{\idn{k}}} {\partial{\idn{d}}}\! \left(\textcolor{olive}{V}\right) G {\textcolor{olive}{V}}^{2} {\textcolor{olive}{X}{\idn{q}}} {\textcolor{olive}{X}{\idn{d}}}}{3}+\frac{2 \kappa  m_{e} {\textcolor{olive}{X}{\idn{q}}} {\partial{\idn{d}}}\! \left(\textcolor{olive}{V}\right) G {\textcolor{olive}{V}}^{2} {\textcolor{olive}{X}{\idn{k}}} {\textcolor{olive}{X}{\idn{d}}}}{3}-\frac{\kappa  m_{e} {\textcolor{olive}{V}}^{3} {\textcolor{olive}{X}{\idn{d}\iup{2}}} {\textcolor{olive}{X}{\idn{k}}} {\partial{\idn{q}}}\! \left(G \right)}{3}+\frac{\kappa  m_{e} {\textcolor{olive}{V}}^{3} {\textcolor{olive}{X}{\idn{d}\iup{2}}} {\textcolor{olive}{X}{\idn{q}}} {\partial{\idn{k}}}\! \left(G \right)}{3}-\frac{\kappa  m_{e} G {\textcolor{olive}{V}}^{3} {\textcolor{olive}{X}{\idn{k}}} {\textcolor{olive}{X}{\idn{q}}}}{3}+\frac{\hslash^{2} {\textcolor{olive}{X}{\idn{k}}} {\partial{\idn{q}}}\! \left(\square \! \left(G \right)\right)}{3}-\frac{\hslash^{2} {\textcolor{olive}{X}{\idn{q}}} {\partial{\idn{k}}}\! \left(\square \! \left(G \right)\right)}{3}+m_{e} \left(\textcolor{olive}{V} {\textcolor{olive}{X}{\idn{k}}} {\partial{\idn{q}}}\! \left(G \right)-\textcolor{olive}{V} {\textcolor{olive}{X}{\idn{q}}} {\partial{\idn{k}}}\! \left(G \right)-\frac{{\textcolor{olive}{X}{\idn{q}}} {\partial{\idn{k}}}\! \left(\textcolor{olive}{V}\right) G}{3}\right) \kappa 
\end{dmath}
\begin{Maple Normal}
Recalling \mapleref{(8)}
{$ \esequiv {\partial{\idn{n}}}(\textcolor{olive}{V})=-{\textcolor{olive}{V}}^{3} {\textcolor{olive}{X}{\idn{n}}} $} and \mapleref{(49)}
{$ \esequiv {\partial{\idn{l}}}(\textcolor{olive}{G})=\frac{\mathrm{i}}{\hslash}{\textcolor{olive}{p}{\idn{l}}} \textcolor{olive}{G} $}
\end{Maple Normal}
\mapleinput
{$ \displaystyle \texttt{>\,} 3\,\hslash \,\mathit{Simplify} (\mathit{SubstituteTensor} (\mapleref{(8)},\mapleref{(49)},\mapleref{(83)})) $}

\begin{dmath}\label{(84)}
\frac{m_{e}^{2} \left[{\textcolor{olive}{Z}{\idn{k}}},{\textcolor{olive}{Z}{\idn{q}}}\right]_{-} G}{\hslash}=-\mathrm{i} \kappa  m_{e} {\textcolor{olive}{V}}^{3} {\textcolor{olive}{X}{\idn{d}\iup{2}}} {\textcolor{olive}{X}{\idn{k}}} {\textcolor{olive}{p}{\idn{q}}} G +\mathrm{i} \kappa  m_{e} {\textcolor{olive}{V}}^{3} {\textcolor{olive}{X}{\idn{d}\iup{2}}} {\textcolor{olive}{X}{\idn{q}}} {\textcolor{olive}{p}{\idn{k}}} G +\hslash^{3} {\textcolor{olive}{X}{\idn{k}}} {\partial{\idn{q}}}\! \left(\square \! \left(G \right)\right)-\hslash^{3} {\textcolor{olive}{X}{\idn{q}}} {\partial{\idn{k}}}\! \left(\square \! \left(G \right)\right)+3 \mathrm{i} m_{e} \kappa  \left(\textcolor{olive}{V} {\textcolor{olive}{X}{\idn{k}}} {\textcolor{olive}{p}{\idn{q}}} G -\textcolor{olive}{V} {\textcolor{olive}{X}{\idn{q}}} {\textcolor{olive}{p}{\idn{k}}} G \right)
\end{dmath}
\begin{Maple Normal}
Evaluating the term 
{$ {\partial{\idn{q}}}(\square (G (X))) $}
\end{Maple Normal}
\mapleinput
{$ \displaystyle \texttt{>\,} p [a]\,p [l]^{2}\cdot G (X) $}

\begin{dmath}\label{(85)}
{\textcolor{olive}{p}{\idn{l}\iup{2}}} {\textcolor{olive}{p}{\idn{a}}} G 
\end{dmath}
\mapleinput
{$ \displaystyle \texttt{>\,} \mapleref{(85)}=\mathit{ApplyProductsOfDifferentialOperators} (\mapleref{(85)}) $}

\begin{dmath}\label{(86)}
{\textcolor{olive}{p}{\idn{l}\iup{2}}} {\textcolor{olive}{p}{\idn{a}}} G =\mathrm{i} \hslash^{3} {\partial{\idn{a}}}\! \left(\square \! \left(G \right)\right)
\end{dmath}
\mapleinput
{$ \displaystyle \texttt{>\,} \mathit{isolate} (\mapleref{(86)},{\partial{\idn{a}}}(\square (G (X)))) $}

\begin{dmath}\label{(87)}
{\partial{\idn{a}}}\! \left(\square \! \left(G \right)\right)=\frac{\mathrm{-i} {\textcolor{olive}{p}{\idn{l}\iup{2}}} {\textcolor{olive}{p}{\idn{a}}} G}{\hslash^{3}}
\end{dmath}
\begin{Maple Normal}
Inserting this result into the expression \mapleref{(84)} for
{$ [{\textcolor{olive}{Z}{\idn{k}}},{\textcolor{olive}{Z}{\idn{q}}}]_{-} $}and removing the test function multiplying by 
{$ G (\textcolor{olive}{X})^{-1} $}
\end{Maple Normal}
\mapleinput
{$ \displaystyle \texttt{>\,} \mathit{Simplify} (\mathit{SubstituteTensor} (\mapleref{(87)},\mapleref{(84)})\cdot G (X)^{\mathit{-}1}) $}

\begin{dmath}\label{(88)}
\frac{m_{e}^{2} \left[{\textcolor{olive}{Z}{\idn{k}}},{\textcolor{olive}{Z}{\idn{q}}}\right]_{-}}{\hslash}=\mathrm{i} \kappa  m_{e} {\textcolor{olive}{V}}^{3} {\textcolor{olive}{X}{\idn{d}\iup{2}}} {\textcolor{olive}{X}{\idn{q}}} {\textcolor{olive}{p}{\idn{k}}}+\mathrm{i} \left(3 \textcolor{olive}{V} {\textcolor{olive}{X}{\idn{k}}} {\textcolor{olive}{p}{\idn{q}}} \kappa  m_{e}-3 \kappa  m_{e} \textcolor{olive}{V} {\textcolor{olive}{X}{\idn{q}}} {\textcolor{olive}{p}{\idn{k}}}-\kappa  m_{e} {\textcolor{olive}{V}}^{3} {\textcolor{olive}{X}{\idn{d}\iup{2}}} {\textcolor{olive}{X}{\idn{k}}} {\textcolor{olive}{p}{\idn{q}}}-{\textcolor{olive}{X}{\idn{k}}} {\textcolor{olive}{p}{\idn{d}\iup{2}}} {\textcolor{olive}{p}{\idn{q}}}+{\textcolor{olive}{X}{\idn{q}}} {\textcolor{olive}{p}{\idn{d}\iup{2}}} {\textcolor{olive}{p}{\idn{k}}}\right)
\end{dmath}
\begin{Maple Normal}
This expression can be factored
\end{Maple Normal}
\mapleinput
{$ \displaystyle \texttt{>\,} \mathit{Factor} (\mapleref{(88)}) $}

\begin{dmath}\label{(89)}
\frac{m_{e}^{2} \left[{\textcolor{olive}{Z}{\idn{k}}},{\textcolor{olive}{Z}{\idn{q}}}\right]_{-}}{\hslash}=\mathrm{-i} \left(-3 \textcolor{olive}{V} \kappa  m_{e}+{\textcolor{olive}{p}{\idn{d}\iup{2}}}+\kappa  m_{e} {\textcolor{olive}{V}}^{3} {\textcolor{olive}{X}{\idn{d}\iup{2}}}\right) \left(-{\textcolor{olive}{X}{\idn{q}}} {\textcolor{olive}{p}{\idn{k}}}+{\textcolor{olive}{X}{\idn{k}}} {\textcolor{olive}{p}{\idn{q}}}\right)
\end{dmath}
\begin{Maple Normal}
Using the identity 
{$ \mapleref{(9)}\esequiv {\textcolor{olive}{V}}^{3} {\textcolor{olive}{X}{\idn{l}}}^{2}=\textcolor{olive}{V} $}for the potential
\end{Maple Normal}
\mapleinput
{$ \displaystyle \texttt{>\,} \frac{\hslash}{\mathit{m_{e}^{2}}}\,\mathit{SubstituteTensor} (\mapleref{(9)},\mapleref{(89)}) $}

\begin{dmath}\label{(90)}
\left[{\textcolor{olive}{Z}{\idn{k}}},{\textcolor{olive}{Z}{\idn{q}}}\right]_{-}=\frac{\mathrm{-i} \hslash  \left(-2 \textcolor{olive}{V} \kappa  m_{e}+{\textcolor{olive}{p}{\idn{d}\iup{2}}}\right) \left(-{\textcolor{olive}{X}{\idn{q}}} {\textcolor{olive}{p}{\idn{k}}}+{\textcolor{olive}{X}{\idn{k}}} {\textcolor{olive}{p}{\idn{q}}}\right)}{m_{e}^{2}}
\end{dmath}
\begin{Maple Normal}
Next using 
\end{Maple Normal}
\mapleinput
{$ \displaystyle \texttt{>\,} \mapleref{(73)},\mapleref{(78)} $}

\begin{dmath}\label{(91)}
{\textcolor{olive}{p}{\idn{l}\iup{2}}} \hiderel{=} 2 \left(\kappa  \textcolor{olive}{V}+\textcolor{olive}{H}\right) m_{e},{\textcolor{olive}{X}{\idn{a}}} {\textcolor{olive}{p}{\idn{b}}}-{\textcolor{olive}{X}{\idn{b}}} {\textcolor{olive}{p}{\idn{a}}} \hiderel{=} {\epsilon{\idn{a}\idn{b}\idn{q}}} {\textcolor{olive}{L}{\idn{q}}}
\end{dmath}
\mapleinput
{$ \displaystyle \texttt{>\,} \mathit{SubstituteTensor} (\mapleref{(91)},\mapleref{(90)}) $}

\begin{dmath}\label{(92)}
\left[{\textcolor{olive}{Z}{\idn{k}}},{\textcolor{olive}{Z}{\idn{q}}}\right]_{-}=\frac{-2 \mathrm{i} \hslash  {\epsilon{\idn{c}\idn{k}\idn{q}}} \textcolor{olive}{H} {\textcolor{olive}{L}{\idn{c}}}}{m_{e}}
\end{dmath}
\mapleinput
{$ \displaystyle \texttt{>\,} \mathit{evalb} (\mapleref{(80)}-\mapleref{(92)}) $}

\begin{dmath}\label{(93)}
\mathit{true} 
\end{dmath}
\begin{Maple Normal}
Which is the expected result. Set now differential operators to \textit{none}.
\end{Maple Normal}
\mapleinput
{$ \displaystyle \texttt{>\,} \mathit{Setup} (\mathit{differentialoperators} =\mathit{none}) $}

\begin{dmath}\label{(94)}
\left[\mathit{differentialoperators} =\mathit{none} \right]
\end{dmath}
\section{\textbf{The square of the norm of the Runge-Lenz vector}}
\begin{Maple Normal}
Taking the square of the definition of 
{$ {\textcolor{olive}{Z}{\idn{k}}} $} and simplifying
\end{Maple Normal}
\mapleinput
{$ \displaystyle \texttt{>\,} \mapleref{(30)}^{2} $}

\begin{dmath}\label{(95)}
{\textcolor{olive}{Z}{\idn{k}\iup{2}}}=\left(\frac{{\epsilon{\idn{a}\idn{b}\idn{k}}} \left({\textcolor{olive}{L}{\idn{a}}} {\textcolor{olive}{p}{\idn{b}}}-{\textcolor{olive}{p}{\idn{a}}} {\textcolor{olive}{L}{\idn{b}}}\right)}{2 m_{e}}+\kappa  \textcolor{olive}{V} {\textcolor{olive}{X}{\idn{k}}}\right) \left(\frac{{\epsilon{\idn{c}\idn{d}\idn{k}}} \left({\textcolor{olive}{L}{\idn{c}}} {\textcolor{olive}{p}{\idn{d}}}-{\textcolor{olive}{p}{\idn{c}}} {\textcolor{olive}{L}{\idn{d}}}\right)}{2 m_{e}}+\kappa  \textcolor{olive}{V} {\textcolor{olive}{X}{\idn{k}}}\right)
\end{dmath}
\mapleinput
{$ \displaystyle \texttt{>\,} 2 m_{e}^{2}\,\mathit{Simplify} (\mapleref{(95)}) $}

\begin{dmath}\label{(96)}
2 m_{e}^{2} {\textcolor{olive}{Z}{\idn{k}\iup{2}}}=2 \kappa  \hslash^{2} {\textcolor{olive}{V}}^{3} {\textcolor{olive}{X}{\idn{a}\iup{2}}} m_{e}-6 \kappa  \hslash^{2} \textcolor{olive}{V} m_{e}+2 \kappa^{2} {\textcolor{olive}{V}}^{2} {\textcolor{olive}{X}{\idn{a}\iup{2}}} m_{e}^{2}-4 {\epsilon{\idn{a}\idn{b}\idn{c}}} {\textcolor{olive}{X}{\idn{a}}} {\textcolor{olive}{p}{\idn{b}}} {\textcolor{olive}{L}{\idn{c}}} \textcolor{olive}{V} \kappa  m_{e}-{\textcolor{olive}{p}{\idn{a}}} {\textcolor{olive}{p}{\idn{b}}} {\textcolor{olive}{L}{\idn{b}}} {\textcolor{olive}{L}{\idn{a}}}+2 {\textcolor{olive}{p}{\idn{a}\iup{2}}} {\textcolor{olive}{L}{\idn{b}\iup{2}}}-{\textcolor{olive}{p}{\idn{a}}} {\textcolor{olive}{L}{\idn{b}}} {\textcolor{olive}{L}{\idn{a}}} {\textcolor{olive}{p}{\idn{b}}}
\end{dmath}
\begin{Maple Normal}
Using the algebraic properties of the potential
\end{Maple Normal}
\mapleinput
{$ \displaystyle \texttt{>\,} \mapleref{(9)},V (X)^{\mathit{-}1}\cdot \mapleref{(9)} $}

\begin{dmath}\label{(97)}
{\textcolor{olive}{V}}^{3} {\textcolor{olive}{X}{\idn{l}\iup{2}}} \hiderel{=} \textcolor{olive}{V},{\textcolor{olive}{V}}^{2} {\textcolor{olive}{X}{\idn{l}\iup{2}}} \hiderel{=} 1
\end{dmath}
\begin{Maple Normal}
the expression \mapleref{(96)} for 
{$ {\textcolor{olive}{Z}{\idn{k}\iup{2}}} $} becomes
\end{Maple Normal}
\mapleinput
{$ \displaystyle \texttt{>\,} \mathit{SubstituteTensor} (\mapleref{(97)},\mapleref{(96)}) $}

\begin{dmath}\label{(98)}
2 m_{e}^{2} {\textcolor{olive}{Z}{\idn{k}\iup{2}}}=-4 \kappa  \hslash^{2} \textcolor{olive}{V} m_{e}+2 \kappa^{2} m_{e}^{2}-4 {\epsilon{\idn{a}\idn{b}\idn{c}}} {\textcolor{olive}{X}{\idn{a}}} {\textcolor{olive}{p}{\idn{b}}} {\textcolor{olive}{L}{\idn{c}}} \textcolor{olive}{V} \kappa  m_{e}-{\textcolor{olive}{p}{\idn{a}}} {\textcolor{olive}{p}{\idn{b}}} {\textcolor{olive}{L}{\idn{b}}} {\textcolor{olive}{L}{\idn{a}}}+2 {\textcolor{olive}{p}{\idn{a}\iup{2}}} {\textcolor{olive}{L}{\idn{b}\iup{2}}}-{\textcolor{olive}{p}{\idn{a}}} {\textcolor{olive}{L}{\idn{b}}} {\textcolor{olive}{L}{\idn{a}}} {\textcolor{olive}{p}{\idn{b}}}
\end{dmath}
\begin{Maple Normal}
The term having 
{$ {\epsilon{\idn{a}\idn{b}\idn{c}}} $} can be simplified using the expression of the momentum operator
\end{Maple Normal}
\mapleinput
{$ \displaystyle \texttt{>\,} (\mathit{rhs} =\mathit{lhs})(\mapleref{(10)}) $}

\begin{dmath}\label{(99)}
{\epsilon{\idn{m}\idn{n}\idn{q}}} {\textcolor{olive}{X}{\idn{m}}} {\textcolor{olive}{p}{\idn{n}}}={\textcolor{olive}{L}{\idn{q}}}
\end{dmath}
\mapleinput
{$ \displaystyle \texttt{>\,} \mapleref{(99)}\cdot L [q]\cdot V (X) $}

\begin{dmath}\label{(100)}
{\epsilon{\idn{m}\idn{n}\idn{q}}} {\textcolor{olive}{X}{\idn{m}}} {\textcolor{olive}{p}{\idn{n}}} {\textcolor{olive}{L}{\idn{q}}} \textcolor{olive}{V}={\textcolor{olive}{L}{\idn{q}\iup{2}}} \textcolor{olive}{V}
\end{dmath}
\mapleinput
{$ \displaystyle \texttt{>\,} \mathit{SubstituteTensor} (\mapleref{(100)},\mapleref{(98)}) $}

\begin{dmath}\label{(101)}
2 m_{e}^{2} {\textcolor{olive}{Z}{\idn{k}\iup{2}}}=-4 \kappa  \hslash^{2} \textcolor{olive}{V} m_{e}+2 \kappa^{2} m_{e}^{2}-4 {\textcolor{olive}{L}{\idn{c}\iup{2}}} \textcolor{olive}{V} \kappa  m_{e}-{\textcolor{olive}{p}{\idn{a}}} {\textcolor{olive}{p}{\idn{b}}} {\textcolor{olive}{L}{\idn{b}}} {\textcolor{olive}{L}{\idn{a}}}+2 {\textcolor{olive}{p}{\idn{a}\iup{2}}} {\textcolor{olive}{L}{\idn{b}\iup{2}}}-{\textcolor{olive}{p}{\idn{a}}} {\textcolor{olive}{L}{\idn{b}}} {\textcolor{olive}{L}{\idn{a}}} {\textcolor{olive}{p}{\idn{b}}}
\end{dmath}
\begin{Maple Normal}
Reordering \mapleref{(101)} to have the two terms with four operators sorted as 
{$ {\textcolor{olive}{p}{\idn{a}}} {\textcolor{olive}{p}{\idn{b}}} {\textcolor{olive}{L}{\idn{a}}} {\textcolor{olive}{L}{\idn{b}}} $}
\end{Maple Normal}
\mapleinput
{$ \displaystyle \texttt{>\,} \mathit{Simplify} (\mathit{SortProducts} (\mapleref{(101)},[p [a],p [b],L [a],L [b]])) $}

\begin{dmath}\label{(102)}
2 m_{e}^{2} {\textcolor{olive}{Z}{\idn{k}\iup{2}}}=-4 \kappa  \hslash^{2} \textcolor{olive}{V} m_{e}+2 \kappa^{2} m_{e}^{2}-4 {\textcolor{olive}{L}{\idn{a}\iup{2}}} \textcolor{olive}{V} \kappa  m_{e}-2 {\textcolor{olive}{p}{\idn{a}}} {\textcolor{olive}{p}{\idn{b}}} {\textcolor{olive}{L}{\idn{a}}} {\textcolor{olive}{L}{\idn{b}}}+2 {\textcolor{olive}{p}{\idn{a}\iup{2}}} {\textcolor{olive}{L}{\idn{b}\iup{2}}}+2 \hslash^{2} {\textcolor{olive}{p}{\idn{a}\iup{2}}}
\end{dmath}
\begin{Maple Normal}
Considering now the resulting single term 
{$ {\textcolor{olive}{p}{\idn{a}}} {\textcolor{olive}{p}{\idn{b}}} {\textcolor{olive}{L}{\idn{a}}} {\textcolor{olive}{L}{\idn{b}}} $}, it can be shown it is equal to zero using the definition \mapleref{(10)}
{$ \esequiv {\textcolor{olive}{L}{\idn{q}}}={\epsilon{\idn{m}\idn{n}\idn{q}}} {\textcolor{olive}{X}{\idn{m}}} {\textcolor{olive}{p}{\idn{n}}} $}
\end{Maple Normal}
\mapleinput
{$ \displaystyle \texttt{>\,} p [a]\,p [b]\,L [a]\,L [b] $}

\begin{dmath}\label{(103)}
{\textcolor{olive}{p}{\idn{a}}} {\textcolor{olive}{p}{\idn{b}}} {\textcolor{olive}{L}{\idn{a}}} {\textcolor{olive}{L}{\idn{b}}}
\end{dmath}
\mapleinput
{$ \displaystyle \texttt{>\,} \mapleref{(103)}=\mathit{SubstituteTensor} (\mapleref{(10)},\mapleref{(103)}) $}

\begin{dmath}\label{(104)}
{\textcolor{olive}{p}{\idn{a}}} {\textcolor{olive}{p}{\idn{b}}} {\textcolor{olive}{L}{\idn{a}}} {\textcolor{olive}{L}{\idn{b}}}={\epsilon{\idn{a}\idn{m}\idn{n}}} {\epsilon{\idn{b}\idn{e}\idn{f}}} {\textcolor{olive}{p}{\idn{a}}} {\textcolor{olive}{p}{\idn{b}}} {\textcolor{olive}{X}{\idn{m}}} {\textcolor{olive}{p}{\idn{n}}} {\textcolor{olive}{X}{\idn{e}}} {\textcolor{olive}{p}{\idn{f}}}
\end{dmath}
\mapleinput
{$ \displaystyle \texttt{>\,} \mathit{Simplify} (\mapleref{(104)}) $}

\begin{dmath}\label{(105)}
{\textcolor{olive}{p}{\idn{a}}} {\textcolor{olive}{p}{\idn{b}}} {\textcolor{olive}{L}{\idn{a}}} {\textcolor{olive}{L}{\idn{b}}}=0
\end{dmath}
\begin{Maple Normal}
Taking this result into account, we have, for 
{$ {\textcolor{olive}{Z}{\idn{k}}}^{2} $}
\end{Maple Normal}
\mapleinput
{$ \displaystyle \texttt{>\,} \mathit{subs} (\mapleref{(105)},\mapleref{(102)}) $}

\begin{dmath}\label{(106)}
2 m_{e}^{2} {\textcolor{olive}{Z}{\idn{k}\iup{2}}}=-4 \kappa  \hslash^{2} \textcolor{olive}{V} m_{e}+2 \kappa^{2} m_{e}^{2}-4 {\textcolor{olive}{L}{\idn{a}\iup{2}}} \textcolor{olive}{V} \kappa  m_{e}+2 {\textcolor{olive}{p}{\idn{a}\iup{2}}} {\textcolor{olive}{L}{\idn{b}\iup{2}}}+2 \hslash^{2} {\textcolor{olive}{p}{\idn{a}\iup{2}}}
\end{dmath}
\begin{Maple Normal}
Substituting now \mapleref{(73)}
{$ \esequiv {\textcolor{olive}{p}{\idn{l}}}^{2}=2 (\kappa  \textcolor{olive}{V}+\textcolor{olive}{H}) m_{e} $}
\end{Maple Normal}
\mapleinput
{$ \displaystyle \texttt{>\,} \frac{1}{\mathit{2\,m_{e}^{2}\,}}\,\mathit{SubstituteTensor} (\mapleref{(73)},\mapleref{(106)}) $}

\begin{dmath}\label{(107)}
{\textcolor{olive}{Z}{\idn{k}\iup{2}}}=\frac{2 \hslash^{2} \textcolor{olive}{H} m_{e}+\kappa^{2} m_{e}^{2}-2 {\textcolor{olive}{L}{\idn{a}\iup{2}}} \textcolor{olive}{V} \kappa  m_{e}+2 \left(\kappa  \textcolor{olive}{V}+\textcolor{olive}{H}\right) m_{e} {\textcolor{olive}{L}{\idn{b}\iup{2}}}}{m_{e}^{2}}
\end{dmath}
\begin{Maple Normal}
Equalizing the repeated indices, the right-hand side can be factored
\end{Maple Normal}
\mapleinput
{$ \displaystyle \texttt{>\,} (\mathit{lhs} =\mathit{Factor} @ \mathit{rhs})(\mathit{EqualizeRepeatedIndices} (\mapleref{(107)})-\kappa^{2})+\kappa^{2} $}

\begin{dmath}\label{(108)}
{\textcolor{olive}{Z}{\idn{k}\iup{2}}}=\frac{2 \textcolor{olive}{H} \left(\hslash^{2}+{\textcolor{olive}{L}{\idn{a}\iup{2}}}\right)}{m_{e}}+\kappa^{2}
\end{dmath}
\begin{Maple Normal}
Which is the result we wanted to demonstrate.
\end{Maple Normal}
\section{\textbf{The atomic hydrogen spectrum}}
\begin{Maple Normal}
We now have all the algebra to reconstruct the hydrogen spectrum. Following the literature, this approach is limited to the bound states for which the energy is negative. Assuming an eigenstate of \textit{H} with negative eigenvalue \textit{E}, we replace the Hamiltonian \textit{H} by \textit{E}, and look for the possible values of \textit{E}. Another way to state the same thing is that the analysis is restricted to the subspace of energy \textit{E}. The operator 
{$ M_{n}=\sqrt{-\frac{m_{e}}{2 E}}\, Z_{n} $}, is introduced as mentioned in Sec.1. The operators \textit{J} and \textit{K}, to be used soon after, are added to the formulation of the problem
\end{Maple Normal}
\mapleinput
{$ \displaystyle \texttt{>\,} \mathit{Setup} (\mathit{hermitianoperators} =\{M ,J ,K \})
\\
  $}

\begin{dmath}\label{(109)}
\left[\mathit{hermitianoperators} =\left\{\textcolor{olive}{H},\textcolor{olive}{J},\textcolor{olive}{K},\textcolor{olive}{L},\textcolor{olive}{M},\textcolor{olive}{V},\textcolor{olive}{p},\textcolor{olive}{x},\textcolor{olive}{y},\textcolor{olive}{z}\right\}\right]
\end{dmath}
\mapleinput
{$ \displaystyle \texttt{>\,} \mathit{Define} (M [n],J [n],K [n],\mathit{quiet}) $}

\begin{dmath}\label{(110)}
\left\{{\textcolor{olive}{\gamma}{\idn{a}}},{\textcolor{olive}{J}{\idn{n}}},{\textcolor{olive}{K}{\idn{n}}},{\textcolor{olive}{L}{\idn{k}}},{\textcolor{olive}{M}{\idn{n}}},{\textcolor{olive}{\sigma}{\idn{a}}},{\textcolor{olive}{Z}{\idn{k}}},{\partial{\idn{a}}},{g{\idn{a}\idn{b}}},{\textcolor{olive}{p}{\idn{k}}},{\epsilon{\idn{a}\idn{b}\idn{c}}},{\textcolor{olive}{X}{\idn{a}}}\right\}
\end{dmath}
\begin{Maple Normal}
The domain for 
{$ m_{e} $} and 
{$ E  $} is set via
\end{Maple Normal}
\mapleinput
{$ \displaystyle \texttt{>\,} \mathit{Assume} (m_{e}>0,E <0) $}

\begin{dmath}\label{(111)}
\left\{E ::\left(-\infty ,0\right)\right\},\left\{m_{e}::\left(0,\infty \right)\right\}
\end{dmath}
\begin{Maple Normal}
from where
\end{Maple Normal}
\mapleinput
{$ \displaystyle \texttt{>\,} M [n]=\sqrt{-\frac{m_{e}}{2 E}} Z [n] $}

\begin{dmath}\label{(112)}
{\textcolor{olive}{M}{\idn{n}}}=\frac{\sqrt{-\frac{2 m_{e}}{E}}\, {\textcolor{olive}{Z}{\idn{n}}}}{2}
\end{dmath}
\mapleinput
{$ \displaystyle \texttt{>\,} \mathit{simplify} (\mathit{isolate} (\mapleref{(112)},Z [n])) $}

\begin{dmath}\label{(113)}
{\textcolor{olive}{Z}{\idn{n}}}=\frac{{\textcolor{olive}{M}{\idn{n}}} \sqrt{2}\, \sqrt{-E}}{\sqrt{m_{e}}}
\end{dmath}
\begin{Maple Normal}
Recalling the commutation rules \mapleref{(92)}
{$ \esequiv [{\textcolor{olive}{Z}{\idn{k}}},{\textcolor{olive}{Z}{\idn{q}}}]_{-}=-\frac{2 \mathrm{i} \hslash  {\epsilon{\idn{a}\idn{k}\idn{q}}} \textcolor{olive}{H} {\textcolor{olive}{L}{\idn{a}}}}{m_{e}} $} and \mapleref{(113)}  above with \textit{E} replacing \textit{H}
\end{Maple Normal}
\mapleinput
{$ \displaystyle \texttt{>\,} \mathit{SubstituteTensor} (H =E ,\mapleref{(113)},\mapleref{(92)}) $}

\begin{dmath}\label{(114)}
\left[\frac{{\textcolor{olive}{M}{\idn{k}}} \sqrt{2}\, \sqrt{-E}}{\sqrt{m_{e}}},\frac{{\textcolor{olive}{M}{\idn{q}}} \sqrt{2}\, \sqrt{-E}}{\sqrt{m_{e}}}\right]_{-}=\frac{-2 \mathrm{i} \hslash  {\epsilon{\idn{c}\idn{k}\idn{q}}} E {\textcolor{olive}{L}{\idn{c}}}}{m_{e}}
\end{dmath}
\mapleinput
{$ \displaystyle \texttt{>\,} \mathit{Simplify} (\mapleref{(114)}) $}

\begin{dmath}\label{(115)}
-\frac{2 E \left[{\textcolor{olive}{M}{\idn{k}}},{\textcolor{olive}{M}{\idn{q}}}\right]_{-}}{m_{e}}=\frac{-2 \mathrm{i} \hslash  {\epsilon{\idn{c}\idn{k}\idn{q}}} E {\textcolor{olive}{L}{\idn{c}}}}{m_{e}}
\end{dmath}
\begin{Maple Normal}
Isolating the commutator, the expression \mapleref{(92)} for 
{$ [{\textcolor{olive}{Z}{\idn{k}}},{\textcolor{olive}{Z}{\idn{q}}}]_{-} $}appears rewritten in terms of the 
{$ {\textcolor{olive}{M}{\idn{k}}} $} as
\end{Maple Normal}
\mapleinput
{$ \displaystyle \texttt{>\,} \mathit{isolate} (\mapleref{(115)},\mathit{Commutator} (M [k],M [q])) $}

\begin{dmath}\label{(116)}
\left[{\textcolor{olive}{M}{\idn{k}}},{\textcolor{olive}{M}{\idn{q}}}\right]_{-}=\mathrm{i} \hslash  {\epsilon{\idn{c}\idn{k}\idn{q}}} {\textcolor{olive}{L}{\idn{c}}}
\end{dmath}
\begin{Maple Normal}
Likewise, inserting \mapleref{(113)}
{$ \esequiv {\textcolor{olive}{Z}{\idn{n}}}=\frac{{\textcolor{olive}{M}{\idn{n}}} \sqrt{2}\, \sqrt{-E}}{\sqrt{m_{e}}} $} into the expression \mapleref{(61)}
{$ \esequiv {\textcolor{gray}{[}{\textcolor{olive}{L}{\idn{q}}},{\textcolor{olive}{Z}{\idn{k}}}\textcolor{gray}{]}}_{-}=-\hslash  \mathrm{i} {\epsilon{\idn{a}\idn{k}\idn{q}}} {\textcolor{olive}{Z}{\idn{a}}} $}, we get it rewritten in terms of 
{$ {\textcolor{olive}{L}{\idn{q}}},{\textcolor{olive}{M}{\idn{k}}} $}
\end{Maple Normal}
\mapleinput
{$ \displaystyle \texttt{>\,} \mathit{Simplify} (\mathit{SubstituteTensor} (\mapleref{(113)},\mapleref{(61)})) $}

\begin{dmath}\label{(117)}
\frac{\sqrt{2}\, \sqrt{-E}\, \left[{\textcolor{olive}{L}{\idn{q}}},{\textcolor{olive}{M}{\idn{k}}}\right]_{-}}{\sqrt{m_{e}}}=\frac{\mathrm{-i} \hslash  {\epsilon{\idn{a}\idn{k}\idn{q}}} {\textcolor{olive}{M}{\idn{a}}} \sqrt{2}\, \sqrt{-E}}{\sqrt{m_{e}}}
\end{dmath}
\mapleinput
{$ \displaystyle \texttt{>\,} \mathit{isolate} (\mapleref{(117)},\mathit{Commutator} (L [q],M [k])) $}

\begin{dmath}\label{(118)}
\left[{\textcolor{olive}{L}{\idn{q}}},{\textcolor{olive}{M}{\idn{k}}}\right]_{-}=\mathrm{-i} \hslash  {\epsilon{\idn{a}\idn{k}\idn{q}}} {\textcolor{olive}{M}{\idn{a}}}
\end{dmath}
\begin{Maple Normal}
Add these two newly derived commutators to the setup
\end{Maple Normal}
\mapleinput
{$ \displaystyle \texttt{>\,} \mathit{Setup} (\mapleref{(116)},\mapleref{(118)}) $}

\begin{dmath}\label{(119)}
\left[\mathit{algebrarules} =\left\{\left[\textcolor{olive}{H},{\textcolor{olive}{L}{\idn{q}}}\right]_{-} \hiderel{=} 0,\left[\textcolor{olive}{H},{\textcolor{olive}{Z}{\idn{k}}}\right]_{-} \hiderel{=} 0,\left[{\textcolor{olive}{L}{\idn{j}}},{\textcolor{olive}{L}{\idn{k}}}\right]_{-} \hiderel{=} \mathrm{i} \hslash  {\epsilon{\idn{j}\idn{k}\idn{n}}} {\textcolor{olive}{L}{\idn{n}}},
\\
\left[{\textcolor{olive}{L}{\idn{q}}},{\textcolor{olive}{M}{\idn{k}}}\right]_{-}\hiderel{=}\mathrm{-i} \hslash  {\epsilon{\idn{a}\idn{k}\idn{q}}} {\textcolor{olive}{M}{\idn{a}}},\left[{\textcolor{olive}{L}{\idn{q}}},{\textcolor{olive}{Z}{\idn{k}}}\right]_{-} \hiderel{=} \mathrm{-i} \hslash  {\epsilon{\idn{a}\idn{k}\idn{q}}} {\textcolor{olive}{Z}{\idn{a}}},\left[{\textcolor{olive}{L}{\idn{q}}},\textcolor{olive}{V}\right]_{-} \hiderel{=} 0,
\\
\left[{\textcolor{olive}{M}{\idn{k}}},{\textcolor{olive}{M}{\idn{q}}}\right]_{-}\hiderel{=}\mathrm{i} \hslash  {\epsilon{\idn{c}\idn{k}\idn{q}}} {\textcolor{olive}{L}{\idn{c}}},\left[{\textcolor{olive}{p}{\idn{j}}},{\textcolor{olive}{L}{\idn{k}}}\right]_{-} \hiderel{=} \mathrm{i} \hslash  {\epsilon{\idn{j}\idn{k}\idn{n}}} {\textcolor{olive}{p}{\idn{n}}},\left[{\textcolor{olive}{p}{\idn{k}}},{\textcolor{olive}{p}{\idn{l}}}\right]_{-} \hiderel{=} 0,
\\
\left[{\textcolor{olive}{p}{\idn{q}}},\textcolor{olive}{V}\right]_{-}\hiderel{=}\mathrm{i} \hslash  {\textcolor{olive}{V}}^{3} {\textcolor{olive}{X}{\idn{q}}},\left[{\textcolor{olive}{p}{\idn{q}}},{\textcolor{olive}{V}}^{3}\right]_{-} \hiderel{=} 3 \mathrm{i} \hslash  {\textcolor{olive}{V}}^{5} {\textcolor{olive}{X}{\idn{q}}},\left[{\textcolor{olive}{X}{\idn{j}}},{\textcolor{olive}{L}{\idn{k}}}\right]_{-} \hiderel{=} \mathrm{i} \hslash  {\epsilon{\idn{j}\idn{k}\idn{n}}} {\textcolor{olive}{X}{\idn{n}}},
\\
\left[{\textcolor{olive}{X}{\idn{k}}},{\textcolor{olive}{p}{\idn{l}}}\right]_{-}\hiderel{=}\mathrm{i} \hslash  {g{\idn{k}\idn{l}}},\left[{\textcolor{olive}{X}{\idn{k}}},\textcolor{olive}{V}\right]_{-} \hiderel{=} 0\right\}\right]
\end{dmath}
\begin{Maple Normal}
These commutators \mapleref{(118)}, \mapleref{(116)}, together with the departing commutator
\end{Maple Normal}
\mapleinput
{$ \displaystyle \texttt{>\,} (\mathit{\%\!Commutator} =\mathit{Commutator})(L [m],L [n]) $}

\begin{dmath}\label{(120)}
\left[{\textcolor{olive}{L}{\idn{m}}},{\textcolor{olive}{L}{\idn{n}}}\right]_{-}=\mathrm{i} \hslash  {\epsilon{\idn{a}\idn{m}\idn{n}}} {\textcolor{olive}{L}{\idn{a}}}
\end{dmath}
\begin{Maple Normal}
constitute a closed form, the algebra of the SO(4) group, that is, the rotation group in dimension 4.
\end{Maple Normal}
\begin{Maple Normal}

\end{Maple Normal}
\begin{Maple Normal}
We now define the two operators \textit{J} and \textit{K} as follows
\end{Maple Normal}
\mapleinput
{$ \displaystyle \texttt{>\,} J [m]=\frac{1}{2}\cdot (L [m]+M [m]) $}

\begin{dmath}\label{(121)}
{\textcolor{olive}{J}{\idn{m}}}=\frac{{\textcolor{olive}{L}{\idn{m}}}}{2}+\frac{{\textcolor{olive}{M}{\idn{m}}}}{2}
\end{dmath}
\mapleinput
{$ \displaystyle \texttt{>\,} K [m]=\frac{1}{2}\cdot (L [m]-M [m]) $}

\begin{dmath}\label{(122)}
{\textcolor{olive}{K}{\idn{m}}}=\frac{{\textcolor{olive}{L}{\idn{m}}}}{2}-\frac{{\textcolor{olive}{M}{\idn{m}}}}{2}
\end{dmath}
\begin{Maple Normal}
Because \textit{M} and \textit{L} both commute with \textit{H} (since \textit{M} is proportional to \textit{Z} up-to a commutative factor), it is straightforward to see that \textit{J} and \textit{K} also commute with \textit{H}\textit{.} They are therefore constants of the motion. Additionally, because at this point (see \mapleref{(119)}) the system already knows about the commutators \mapleref{(116)}
{$ \esequiv [{\textcolor{olive}{M}{\idn{k}}},{\textcolor{olive}{M}{\idn{q}}}]_{-} $} and \mapleref{(118)}
{$ \esequiv [{\textcolor{olive}{L}{\idn{q}}},{\textcolor{olive}{M}{\idn{k}}}]_{-} $}, the commutator between the components of 
{$ {\textcolor{olive}{J}{\idn{m}}} $} results in
\end{Maple Normal}
\mapleinput
{$ \displaystyle \texttt{>\,} \mathit{Commutator} (\mapleref{(121)},\mathit{SubstituteTensorIndices} (m =n ,\mapleref{(121)})) $}

\begin{dmath}\label{(123)}
\left[{\textcolor{olive}{J}{\idn{m}}},{\textcolor{olive}{J}{\idn{n}}}\right]_{-}=\frac{\mathrm{i}}{4} \left(\left({\textcolor{olive}{L}{\idn{a}}}+2 {\textcolor{olive}{M}{\idn{a}}}\right) {\epsilon{\idn{a}\idn{m}\idn{n}}}+{\epsilon{\idn{c}\idn{m}\idn{n}}} {\textcolor{olive}{L}{\idn{c}}}\right) \hslash 
\end{dmath}
\mapleinput
{$ \displaystyle \texttt{>\,} \mathit{Simplify} (\mapleref{(123)}) $}

\begin{dmath}\label{(124)}
\left[{\textcolor{olive}{J}{\idn{m}}},{\textcolor{olive}{J}{\idn{n}}}\right]_{-}=\frac{\mathrm{i}}{2} {\epsilon{\idn{a}\idn{m}\idn{n}}} \hslash  \left({\textcolor{olive}{L}{\idn{a}}}+{\textcolor{olive}{M}{\idn{a}}}\right)
\end{dmath}
\mapleinput
{$ \displaystyle \texttt{>\,} \mathit{SubstituteTensor} ((\mathit{rhs} =\mathit{lhs})(\mapleref{(121)}),\mapleref{(124)}) $}

\begin{dmath}\label{(125)}
\left[{\textcolor{olive}{J}{\idn{m}}},{\textcolor{olive}{J}{\idn{n}}}\right]_{-}=\mathrm{i} {\epsilon{\idn{a}\idn{m}\idn{n}}} \hslash  {\textcolor{olive}{J}{\idn{a}}}
\end{dmath}
\begin{Maple Normal}
In a similar manner
\end{Maple Normal}
\mapleinput
{$ \displaystyle \texttt{>\,} \mathit{Commutator} (\mapleref{(122)},\mathit{SubstituteTensorIndices} (m =n ,\mapleref{(122)})) $}

\begin{dmath}\label{(126)}
\left[{\textcolor{olive}{K}{\idn{m}}},{\textcolor{olive}{K}{\idn{n}}}\right]_{-}=\frac{\mathrm{i}}{4} \left(\left({\textcolor{olive}{L}{\idn{a}}}-2 {\textcolor{olive}{M}{\idn{a}}}\right) {\epsilon{\idn{a}\idn{m}\idn{n}}}+{\epsilon{\idn{c}\idn{m}\idn{n}}} {\textcolor{olive}{L}{\idn{c}}}\right) \hslash 
\end{dmath}
\mapleinput
{$ \displaystyle \texttt{>\,} \mathit{Simplify} (\mapleref{(126)}) $}

\begin{dmath}\label{(127)}
\left[{\textcolor{olive}{K}{\idn{m}}},{\textcolor{olive}{K}{\idn{n}}}\right]_{-}=\frac{\mathrm{i}}{2} {\epsilon{\idn{a}\idn{m}\idn{n}}} \hslash  \left({\textcolor{olive}{L}{\idn{a}}}-{\textcolor{olive}{M}{\idn{a}}}\right)
\end{dmath}
\mapleinput
{$ \displaystyle \texttt{>\,} \mathit{SubstituteTensor} ((\mathit{rhs} =\mathit{lhs})(\mapleref{(122)}),\mapleref{(127)}) $}

\begin{dmath}\label{(128)}
\left[{\textcolor{olive}{K}{\idn{m}}},{\textcolor{olive}{K}{\idn{n}}}\right]_{-}=\mathrm{i} {\epsilon{\idn{a}\idn{m}\idn{n}}} \hslash  {\textcolor{olive}{K}{\idn{a}}}
\end{dmath}
\begin{Maple Normal}
Also
\end{Maple Normal}
\mapleinput
{$ \displaystyle \texttt{>\,} \mathit{Commutator} (\mapleref{(121)},\mathit{subs} (m =n ,\mapleref{(122)})) $}

\begin{dmath}\label{(129)}
\left[{\textcolor{olive}{J}{\idn{m}}},{\textcolor{olive}{K}{\idn{n}}}\right]_{-}=\frac{\mathrm{i}}{4} \hslash  \left({\epsilon{\idn{a}\idn{m}\idn{n}}} {\textcolor{olive}{L}{\idn{a}}}-{\epsilon{\idn{c}\idn{m}\idn{n}}} {\textcolor{olive}{L}{\idn{c}}}\right)
\end{dmath}
\mapleinput
{$ \displaystyle \texttt{>\,} \mathit{Simplify} (\mapleref{(129)}) $}

\begin{dmath}\label{(130)}
\left[{\textcolor{olive}{J}{\idn{m}}},{\textcolor{olive}{K}{\idn{n}}}\right]_{-}=0
\end{dmath}
\begin{Maple Normal}
Both \textit{J} and \textit{K} have the symmetry of a rotation operator in two independent 3 dimension spaces. \textit{H} then has the symmetry of the group 
{$ {\mathrm{SO}(3)\otimes \mathrm{SO}(3)} $}. Furthermore, one knows that the possible eigenvalues for the rotation operators \textit{J} and \textit{K} are 
{$ j (j +1)\hslash^{2} $} and 
{$ k (k +1)\hslash^{2} $}, with 
{$ j ,k \in {\{0,\frac{1}{2},1,\frac{3}{2},2,...\}} $}. Computing now 
{$ {\textcolor{olive}{J}}^{2} $}
\end{Maple Normal}
\mapleinput
{$ \displaystyle \texttt{>\,} \mathit{Expand} (\mapleref{(121)}^{2}) $}

\begin{dmath}\label{(131)}
{\textcolor{olive}{J}{\idn{m}\iup{2}}}=\frac{{\textcolor{olive}{L}{\idn{m}\iup{2}}}}{4}+\frac{{\textcolor{olive}{L}{\idn{m}}} {\textcolor{olive}{M}{\idn{m}}}}{2}+\frac{{\textcolor{olive}{M}{\idn{m}\iup{2}}}}{4}
\end{dmath}
\begin{Maple Normal}
Recalling \mapleref{(66)}
{$ \esequiv {\textcolor{olive}{L}{\idn{k}}} {\textcolor{olive}{Z}{\idn{k}}}=0 $}, and considering that \textit{M} is proportional to \textit{Z}, we have that 
{$ {\textcolor{olive}{L}{\idn{m}}} {\textcolor{olive}{M}{\idn{m}}}=0 $}
\end{Maple Normal}
\mapleinput
{$ \displaystyle \texttt{>\,} \mathit{subs} (L [m]\,M [m]=0,\mapleref{(131)}) $}

\begin{dmath}\label{(132)}
{\textcolor{olive}{J}{\idn{m}\iup{2}}}=\frac{{\textcolor{olive}{L}{\idn{m}\iup{2}}}}{4}+\frac{{\textcolor{olive}{M}{\idn{m}\iup{2}}}}{4}
\end{dmath}
\begin{Maple Normal}
Likewise, for 
{$ {\textcolor{olive}{K}}^{2} $}, from \mapleref{(122)}
{$ \esequiv {\textcolor{olive}{K}{\idn{m}}}=\frac{{\textcolor{olive}{L}{\idn{m}}}}{2}-\frac{{\textcolor{olive}{M}{\idn{m}}}}{2} $}
\end{Maple Normal}
\mapleinput
{$ \displaystyle \texttt{>\,} \mathit{Expand} (\mapleref{(122)}^{2}) $}

\begin{dmath}\label{(133)}
{\textcolor{olive}{K}{\idn{m}\iup{2}}}=\frac{{\textcolor{olive}{L}{\idn{m}\iup{2}}}}{4}-\frac{{\textcolor{olive}{L}{\idn{m}}} {\textcolor{olive}{M}{\idn{m}}}}{2}+\frac{{\textcolor{olive}{M}{\idn{m}\iup{2}}}}{4}
\end{dmath}
\mapleinput
{$ \displaystyle \texttt{>\,} \mathit{subs} (L [m]\,M [m]=0,\mapleref{(133)}) $}

\begin{dmath}\label{(134)}
{\textcolor{olive}{K}{\idn{m}\iup{2}}}=\frac{{\textcolor{olive}{L}{\idn{m}\iup{2}}}}{4}+\frac{{\textcolor{olive}{M}{\idn{m}\iup{2}}}}{4}
\end{dmath}
\begin{Maple Normal}
So that 
\end{Maple Normal}
\mapleinput
{$ \displaystyle \texttt{>\,} \mapleref{(132)}-\mapleref{(134)} $}

\begin{dmath}\label{(135)}
{\textcolor{olive}{J}{\idn{m}\iup{2}}}-{\textcolor{olive}{K}{\idn{m}\iup{2}}}=0
\end{dmath}
\begin{Maple Normal}
That is, 
{$ {\textcolor{olive}{J}_{m}}^{2}={\textcolor{olive}{K}_{m}}^{2}, $} which means they share the same eigenvalues, say 
{$ j (j +1)\hslash^{2} $} for a given eigenstate of \textit{H} with the considered eigenvalue \textit{E}. Next, inserting \mapleref{(113)}
{$ \esequiv {\textcolor{olive}{Z}{\idn{n}}}=\frac{{\textcolor{olive}{M}{\idn{n}}} \sqrt{2}\, \sqrt{-E}}{\sqrt{m_{e}}} $} into  \mapleref{(108)}
{$ \esequiv {\textcolor{olive}{Z}{\idn{k}\iup{2}}}=\frac{2 \textcolor{olive}{H} (\hslash^{2}+{\textcolor{olive}{L}{\idn{a}\iup{2}}})}{m_{e}}+\kappa^{2} $} we get an expression for 
{$ {\textcolor{olive}{M}{\idn{k}\iup{2}}} $} 
\end{Maple Normal}
\mapleinput
{$ \displaystyle \texttt{>\,} \mathit{SubstituteTensor} (H =E ,\mapleref{(113)},\mapleref{(108)}) $}

\begin{dmath}\label{(136)}
-\frac{2 E {\textcolor{olive}{M}{\idn{k}\iup{2}}}}{m_{e}}=\frac{2 E \left(\hslash^{2}+{\textcolor{olive}{L}{\idn{a}\iup{2}}}\right)}{m_{e}}+\kappa^{2}
\end{dmath}
\mapleinput
{$ \displaystyle \texttt{>\,} -\frac{m_{e}}{2\,E}\,\mapleref{(136)} $}

\begin{dmath}\label{(137)}
{\textcolor{olive}{M}{\idn{k}\iup{2}}}=-\frac{2 \hslash^{2} E +2 E {\textcolor{olive}{L}{\idn{a}\iup{2}}}+\kappa^{2} m_{e}}{2 E}
\end{dmath}
\begin{Maple Normal}
Substituting this result into \mapleref{(132)}
{$ \esequiv  $}
{$ {\textcolor{olive}{J}{\idn{m}\iup{2}}}=\frac{{\textcolor{olive}{L}{\idn{m}\iup{2}}}}{4}+\frac{{\textcolor{olive}{M}{\idn{m}\iup{2}}}}{4} $} and simplifying we get
\end{Maple Normal}
\mapleinput
{$ \displaystyle \texttt{>\,} \mathit{Simplify} (\mathit{SubstituteTensor} (\mapleref{(137)},\mapleref{(132)})) $}

\begin{dmath}\label{(138)}
{\textcolor{olive}{J}{\idn{m}\iup{2}}}=-\frac{\hslash^{2}}{4}-\frac{\kappa^{2} m_{e}}{8 E}
\end{dmath}
\begin{Maple Normal}
Taking the average value of 
{$ {\textcolor{olive}{J}{\idn{m}\iup{2}}} $} over an eigenvector, 
{$ {\textcolor{olive}{J}{\idn{m}\iup{2}}} $} can be replaced by its eigenvalue 
{$ j (j +1)\hslash^{2} $}
\end{Maple Normal}
\mapleinput
{$ \displaystyle \texttt{>\,} \mathit{subs} (J [m]^{2}=j \,(j +1)\,\hslash^{2},\mapleref{(138)}) $}

\begin{dmath}\label{(139)}
j \left(j +1\right) \hslash^{2}=-\frac{\hslash^{2}}{4}-\frac{\kappa^{2} m_{e}}{8 E}
\end{dmath}
\begin{Maple Normal}
from where the possible values of the energy are
\end{Maple Normal}
\mapleinput
{$ \displaystyle \texttt{>\,} \mathit{isolate} (\mapleref{(139)},E) $}

\begin{dmath}\label{(140)}
E =-\frac{\kappa^{2} m_{e}}{2 \hslash^{2} \left(2 j +1\right)^{2}}
\end{dmath}
\begin{Maple Normal}
Assuming 
{$ n =2 j +1 $}, a positive integer and 
{$ j \in {\{0,\frac{1}{2},1,\frac{3}{2},2,...\}}, $}the spectrum for an hydrogen atom is thus
\end{Maple Normal}
\mapleinput
{$ \displaystyle \texttt{>\,} \mathit{subs} (\{2\,j +1=n ,E =E (n)\},\mapleref{(140)}) $}

\begin{dmath}\label{(141)}
E \! \left(n \right)=-\frac{\kappa^{2} m_{e}}{2 \hslash^{2} n^{2}}
\end{dmath}
\begin{Maple Normal}
Which is the energy spectrum for a spinless hydrogenoid system.
\end{Maple Normal}
\section*{\textbf{Conclusions}}
\begin{Maple Normal}
In this presentation, we derived, step-by-step, the SO(4) symmetry of the hydrogen atom and its spectrum using the computer algebra Maple system. The derivation was performed without departing from the results, entering only the main definition formulas in eqs. \mapleref{(1)}, \mapleref{(2)} and \mapleref{(5)}, followed by using a few simplification commands - mainly \textit{Simplify}, \textit{SortProducts} and \textit{SubstituteTensor }- and a handful of Maple basic commands, \textit{subs}, \textit{lhs}, \textit{rhs} and \textit{isolate}. The computational path that was used to get the results of sections 2 to 8 is not unique. Instead of searching for the shortest path, we prioritized clarity and illustration of the techniques that can be used to work on problems like this one. \

\

This problem is mainly about simplifying expressions using two different techniques. First, expressions with noncommutative operands in products need reduction with respect to the commutator algebra rules that have been set. Second, products of tensorial operators require simplification using the sum rule for repeated indices and the symmetries of tensorial subexpressions. Those techniques, which are part of the Maple Physics simplifier, together with the \textit{SortProducts} and \textit{SubstituteTensor} commands for sorting the operands in products to apply tensorial identities, sufficed. The derivations were performed in a reasonably small number of steps. \

\

Two different computational strategies - with and without differential operators - were used in sections 3 and 5, showing a way to verify results, a relevant issue in general when performing complicated algebraic manipulations. The Maple Physics' ability to handle differential operators as noncommutative operands in products (as frequently done in paper and pencil computations) facilitates readability and ease in entering the computations. The complexity of those operations is then handled by one \textit{Physics:-Library} command, \textit{ApplyProductsOfDifferentialOperators} (see eqs. \mapleref{(45)} and \mapleref{(82)}).\

\

It is interesting to note: a) the ability of the system to factor expressions involving products of noncommutative operands (see eqs. \mapleref{(89)} and \mapleref{(108)}) and b) the adaptation of the algorithms for simplifying tensorial expressions [6] to the noncommutativity domain, used throughout this presentation.\

\

Also worth mentioning, the use of equation labels can reduce the whole computation to entering the main definitions, followed by applying a few commands to equation labels. That approach helps to reduce the chance of typographical errors to a very strict minimum. Likewise, the fact that commands and equations distribute over each other allows cumbersome manipulations to be performed in simple ways, as done, for instance, in eqs. \mapleref{(8)}, \mapleref{(9)} and \mapleref{(13)}. Additionally , it was helpful to have the display of each intermediate result automatically expressed using standard mathematical physics notation.
\end{Maple Normal}
\begin{Maple Normal}

\end{Maple Normal}
\begin{Maple Normal}
Finally, if this work focused on a well-known case, the employed tools can be used to tackle a wide range of hot topics research in the quantum mechanics field, and beyond. To give but a few examples, recently (work in progress - see related Mapleprimes post [7]), we reproduced the calculus performed in [8]. This paper evaluates the constraints of magnetostatic traps for neutral cold atoms and Bose Einstein condensates. Besides, the overall possibilities includes the general framework of periodically driven systems, notably requiring Taylor development to approximate commutators [9]. The calculus could be extended to Lie superalgebra in the field of Anderson localization for disordered media, see [10] and supplemental material. Note that the present calculus are performed using an Euclidean metric. This feature could however easily be extended to any arbitrary metric, openings-up a wide range of possibilities to what can now be done with a computer, replacing pencil and paper.
\end{Maple Normal}
\section{\textbf{Acknowledgements}}
\begin{Maple Normal}
This work was partly supported by the French Ministry of Higher Education and Research, Hauts-de-France Regional Council and European Regional Development Fund (ERDF) through the Contrat de Projets Etat-Region (CPER Photonics for Society, P4S).
\end{Maple Normal}
\section*{\textbf{Appendix}}
\begin{Maple Normal}
In this presentation, the input lines are preceded by a prompt
{$ > $} and the commands used are of three kinds: some basic Maple manipulation commands, the main Physics package commands to set the context of a formulation and simplify expressions, and two commands of the \textit{Physics:-Library} to perform specialized operations in expressions.
\end{Maple Normal}

\begin{Maple Normal}
\end{Maple Normal}
\begin{Maple Normal}
\textbf{The basic Maple commands used}
\end{Maple Normal}
\begin{Maple Bullet Item}
\item \textit{interface} is used once at the beginning to set the letter used to represent the imaginary unit (default is I but we used i).
\item \textit{isolate} is used in several places to isolate a variable in an expression, for example isolating \textit{x} in 
{$ a x +b =0 $} results in 
{$ x =-\frac{b}{a} $}
\item \textit{lhs} and \textit{rhs} respectively get the left-hand side 
{$ A  $}and right-hand side 
{$ B  $}of an equation 
{$ A =B  $}
\item \textit{subs} substitutes the left-hand side of an equation by the right-hand side in a given target, for example 
{$ \mathit{subs} (A =B ,A +C) $} results in 
{$ B +C  $}
\item @ is used to compose commands. So
{$ (A \circ B)(x) $} is the same as 
{$ A (B (x)) $}. This command is useful to express an abstract combo of manipulations, for example as in \mapleref{(108)}
{$ \esequiv (\mathit{lhs} =\mathit{Factor} \circ \mathit{rhs}) $}.
\end{Maple Bullet Item}
\begin{Maple Normal}
\textbf{The Physics commands used}
\end{Maple Normal}
\begin{Maple Bullet Item}
\item \textit{Setup} is used to set algebra rules as well as the dimension of space, type of metric, and conventions as the kind of letter used to represent indices.
\item \textit{Commutator} computes the commutator between two objects using the algebra rules set using \textit{Setup}. If no rules are known to the system, it outputs a representation for the commutator that the system understands.
\item \textit{CompactDisplay} is used to avoid redundant display of the functionality of a function.
\item The input 
{$ \textit{d\_} [n] $} represents the 
{$ {\partial{\idn{n}}} $} tensorial differential operator.
\item \textit{Define }is used to define tensors, with or without specifying its components.
\item \textit{Dagger}  computes the Hermitian transpose of an expression.
\item \textit{Normal, Expand, Factor} respectively normalizes, expands and factorizes expressions that involve products of noncommutative operands.
\item \textit{Simplify} performs simplification of tensorial expressions involving products of noncommutative factors taking into account Einstein's sum rule for repeated indices, symmetries of the indices of tensorial subexpressions and custom commutator algebra rules.
\item \textit{SortProducts} uses the commutation rules set using \textit{Setup} to sort the non-commutative operands of a product in an indicated ordering.
\end{Maple Bullet Item}
\begin{Maple Normal}
\textbf{The Physics:-Library commands used}
\end{Maple Normal}
\begin{Maple Bullet Item}
\item \textit{Library:-ApplyProductsOfDifferentialOperators} applies the differential operators found in a product to the product operands that appear to its right. For example, applying this command to  
{$ p V (X) m_{e} $} results in 
{$ m_{e}\cdot p (V (X)) $}
\item \textit{Library:-EqualizeRepeatedIndices} equalizes the repeated indices in the terms of a sum, so for instance applying this command to 
{$ {\textcolor{olive}{L}{\idn{a}\iup{2}}}+{\textcolor{olive}{L}{\idn{b}\iup{2}}} $} results in 
{$ 2\cdot {\textcolor{olive}{L}{\idn{a}\iup{2}}} $}
\end{Maple Bullet Item}

\section*{\textbf{References}}

[1] W. Pauli, "On the hydrogen spectrum from the standpoint of the new quantum mechanics,” Z. Phys. \textbf{36}, 336–363 (1926)
\begin{Maple Normal}

\end{Maple Normal}
[2] S. Weinberg, "Lectures on Quantum Mechanics, second edition," Cambridge University Press, 2015.
\begin{Maple Normal}

\end{Maple Normal}
[3] R. Gilmore, "Lie Groups, Physics, and Geometry: An Introduction for Physicists, Engineers and Chemists," Cambridge University Press, 2008.
\begin{Maple Normal}

\end{Maple Normal}
[4] Veronika Gáliková, Samuel Kováčik, and Peter Prešnajder, "Laplace-Runge-Lenz vector in quantum mechanics in noncommutative space", J. Math. Phys. \textbf{54}, 122106 (2013)
\begin{Maple Normal}

\end{Maple Normal}
[5] Castro, P.G., Kullock, R. "Physics of the
{$ \mathit{so}_{q}(4) $} hydrogen atom". Theor. Math. Phys. \textbf{185}, 1678–1684 (2015).
\begin{Maple Normal}

\end{Maple Normal}
[6] L. R. U. Manssur, R. Portugal, and B. F. Svaiter, "Group-Theoretic Approach for Symbolic Tensor Manipulation," International Journal of Modern Physics C \textbf{13}, 859-879 (2002).
\begin{Maple Normal}

\end{Maple Normal}
[7] P. Szriftgiser, E. S. Cheb-Terrab, \href{https://www.mapleprimes.com/posts/208635-Magnetic-Traps-In-Coldatom-Physics}{Magnetic traps in cold-atom physics} (2017).
\begin{Maple Normal}

\end{Maple Normal}
[8] R. Gerritsma, R. J. C. Spreeuw, "Topological constraints on magnetostatic traps," Phys. Rev. A \textbf{74},  043405 (2006).
\begin{Maple Normal}

\end{Maple Normal}
[9] N. Goldman and J. Dalibard, "Periodically Driven Quantum Systems: Effective Hamiltonians and Engineered Gauge Fields," Phys. Rev. X \textbf{4}, 031027 (2014).
\begin{Maple Normal}

\end{Maple Normal}
[10] E. Khalaf, P. M. Ostrovsky, "Localization Effects on Magnetotransport of a Disordered Weyl Semimetal," Phys. Rev. Lett. \textbf{119}, 106601 (2017).

\end{document}